\documentclass[journal]{IEEEtran}
\usepackage{cite}
\usepackage{amsmath,amssymb,amsfonts}
\usepackage{algorithmic}
\usepackage{graphicx}
\usepackage{textcomp}
\usepackage{multirow,multicol}
\usepackage{comment}
\usepackage[table,xcdraw]{xcolor}
\usepackage{hhline}

\ifCLASSOPTIONcompsoc
\usepackage[caption=false,font=normalsize,labelfon
t=sf,textfont=sf]{subfig}
\else
\usepackage[caption=false,font=footnotesize]{subfig}
\fi

\begin{document}
\title{Pruning for Improved ADC Efficiency in Crossbar-based Analog In-memory Accelerators}

\author{\IEEEauthorblockN{Timur Ibrayev, Isha Garg, Indranil Chakraborty, and Kaushik Roy}\\
\IEEEauthorblockA{School of Electrical and Computer Engineering, Purdue University\\
West Lafayette, USA\\
Emails: (tibrayev, gargi, ichakra, kaushik)@purdue.edu
}
\thanks{The research was funded in part by C-BRIC, one of six centers in JUMP, a Semiconductor Research Corporation (SRC) program sponsored by DARPA, the National Science Foundation, Sandia National Laboratories, Intel Corporation and Vannevar Bush Faculty Fellowship.}}

\markboth
{Pruning for Improved ADC Efficiency in Crossbar-based Analog In-memory Accelerators}
{Pruning for Improved ADC Efficiency in Crossbar-based Analog In-memory Accelerators}

\maketitle

\begin{abstract}
Deep learning has proved successful in many applications but suffers from high computational demands and requires custom accelerators for deployment. Crossbar-based analog in-memory architectures are attractive for acceleration of deep neural networks (DNN), due to their high data reuse and high efficiency enabled by combining storage and computation in memory. However, they require analog-to-digital converters (ADCs) to communicate crossbar outputs. ADCs consume a significant portion of energy and area of every crossbar processing unit, thus diminishing the potential efficiency benefits. Pruning is a well-studied technique to improve the efficiency of DNNs but requires modifications to be effective for crossbars. 
In this paper, we motivate crossbar-attuned pruning to target ADC-specific inefficiencies. This is achieved by identifying three key properties (dubbed \textbf{D.U.B.}) that induce sparsity that can be utilized to reduce ADC energy without sacrificing accuracy. The first property ensures that sparsity translates effectively to hardware efficiency by restricting sparsity levels to \textbf{D}iscrete powers of 2. The other 2 properties encourage columns in the same crossbar to achieve both \textbf{U}nstructured and \textbf{B}alanced sparsity in order to amortize the accuracy drop. 
The desired \textbf{D.U.B.} sparsity is then achieved by regularizing the variance of $L_{0}$ norms of neighboring columns within the same crossbar. 
Our proposed implementation allows it to be directly used in end-to-end gradient-based training.
We apply the proposed algorithm to convolutional layers of VGG11 and ResNet18 models, trained on CIFAR-10 and ImageNet datasets, and achieve up to $7.13\times$ and $1.27\times$ improvement, respectively, in ADC energy with less than 1\% drop in accuracy.
\end{abstract}


\section{Introduction}
\label{sec:introduction}
The rapid advancements in Deep Neural Networks (DNNs) have resulted in their ubiquitous use and incredible performance on many tasks, such as visual and natural language processing~\cite{simonyan2014very, he2016deep, pmlr-v202-frantar23a, zong2023detrs}. 
However, the overparameterization of neural networks demands increased computational and storage requirements~\cite{han2015deep}.
To tackle these issues, different hardware solutions have been proposed in the form of domain-specific accelerators~\cite{akopyan2015truenorth, shafiee2016isaac, chi2016prime, song2017pipelayer, ankit2019puma, ankit2020panther, wan2022compute, yang2019sparse, chen202115, ali202135, kim2023samba, ogbogu2023energy}.

These works leverage the high dependence of neural networks on linear and convolutional operations by converting and performing both these operations in the form of more efficient matrix-vector multiplication (MVM).

Resistive crossbar-based analog in-memory computing offers an attractive way to improve DNN inference by implementing such highly efficient MVM units, which improve efficiency by combining storage and processing elements~\cite{roy2020memory}. 
By storing each DNN weight as the value of a multi-state device at each cross point of the \textit{crossbar array}, the crossbar-based mapping thus, offers high storage density and enables high data reuse. Applying inputs as analog signals to the rows of the crossbar performs an \textit{in-place analog dot product} with stored weights, enabling highly efficient parallel in-situ MVM computations.

\begin{figure*}[ht!]
    \centering
    \includegraphics[width=0.85\linewidth]{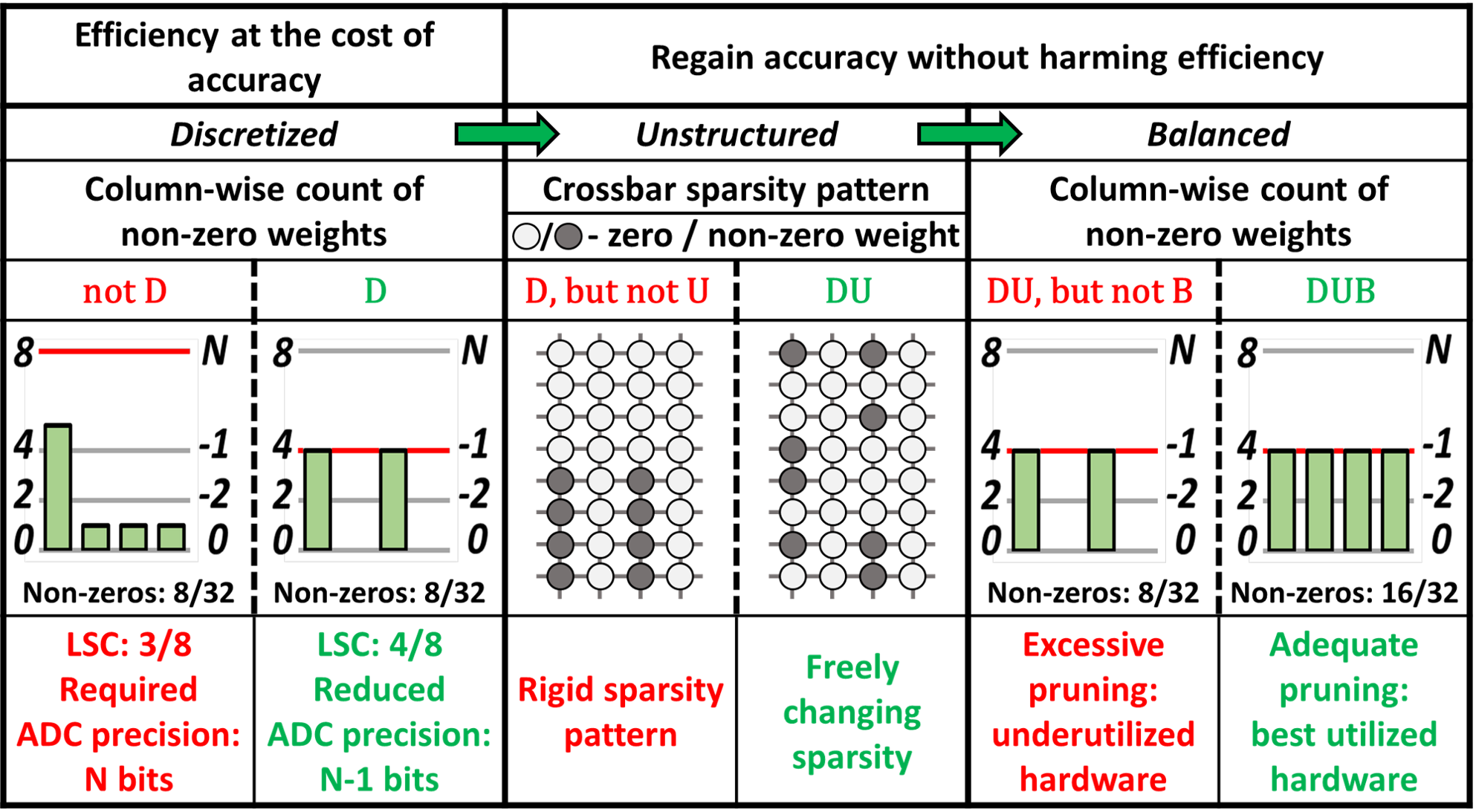}
    \vspace{-3.0mm}
    \caption{Intuition behind discretized, unstructured, and balanced (D.U.B.) properties illustrated through column-wise weight distributions and sparsity patterns of $8\times4$ crossbar by default requiring ADC with precision of $N$ bits. Here, LSC denotes the sparsity (number of zeros) of the least sparse column.}
    \vspace{-3.0mm}
    \label{fig:sparsity}
\end{figure*}

However, analog computations pose considerable challenges, such as having a high peripheral overheads stemming from the need to use ADCs~\cite{roy2020memory, chakraborty2020resistive, huang2021mixed}.
This is because, while ADCs are essential to communicate across multiple crossbar-based analog primitives, they can constitute up to 80\% of total energy and 70\% of the processing core area~\cite{shafiee2016isaac}.
This diminishes the potential efficiency benefits offered by the crossbars and necessitates design techniques that target ADC-specific inefficiencies.

A well studied way of improving efficiency is pruning~\cite{blalock2020what}, wherein insignificant weights are zeroed out. 
For traditional hardware, such as GPUs, pruning methods were proposed to induce unstructured sparsity (individual weights in DNNs)~\cite{han2015deep, han2015learning, zhang2018systematic, rathi2018stdp} or structured sparsity (filters, channels, layers in DNNs)~\cite{wen2016learning, li2016pruning, garg2019low, yang2020harmonious, aketi2020gradual}.
Hardware solutions for crossbar-based architectures were proposed to make use of the resulting sparse DNNs~\cite{yang2019sparse, chen202115, ali202135, kim2023samba, ogbogu2023energy}. However, the required substantial changes to the architecture components necessitate algorithmic approaches inducing sparsity specifically for crossbars.
Consequently, various structured pruning methods were proposed for crossbar structures, such as crossbar rows~\cite{chu2020pim}, crossbar columns~\cite{liang2018crossbar, lin2019learning, chu2020pim}, or entire crossbars~\cite{liang2018crossbar, ankit2019trannsformer, chu2020pim}.
However, these methods mainly aim at reducing the number of crossbars required to map trained DNNs, which does not directly address ADC-specific inefficiencies.
Moreover, such pruning requires modifications to the underlying mapping scheme (i.e. determining weights that are grouped together to a crossbar) and overlooks the plausibility of utilizing unstructured sparsity patterns.

In this work, we propose a pruning method that improves ADC efficiency in crossbar-based accelerators while maintaining accuracy. 
The proposed method results in sparse DNN with three key properties that work together to improve both accuracy and efficiency.
We abbreviate these as the \textbf{D, U, and B} properties. 
The \textbf{D} property focuses on efficiency at the cost of accuracy and the \textbf{U} and \textbf{B} properties work to amortize the drop in accuracy without harming efficiency as shown in Fig.~\ref{fig:sparsity}.

The ADC energy is dictated by its bit precision. 
If the same ADC is re-utilized by all the individual columns of a crossbar for efficiency, the bit precision requirement of the ADC depends only on the sparsity of the least sparse column in the crossbar. 
To ensure that pruning-induced sparsity translates to reduced ADC bit-precision, we restrict the allowed sparsity levels of the least sparse column to powers of 2. 
We term this the \textbf{Discretization} property (\textbf{D}) and illustrate it by weight distributions shown in the two leftmost columns of Fig.~\ref{fig:sparsity}.

The \textbf{D} property defines the base sparsity level of the crossbar but does not enforce any constraints on the pruning patterns. 
If the sparsity of the least sparse column in a crossbar achieves any of the discretized levels, the locations of zeros in a column are irrelevant.
As a result, unstructured sparsity is more favorable than rigidly structured sparsity, as it attains higher compression ratios with lower accuracy drops due to the freedom in sparsity patterns.
Hence, the lack of structural constraints on DNN weights provides a way to amortize the drop in accuracy resulting from the aggressive removal of weights encouraged by \textbf{D} property. 
This is termed as the \textbf{Unstructured} property (\textbf{U}) and can be observed from two crossbar sparsity patterns shown in Fig.~\ref{fig:sparsity}.

After lifting the restrictions on the \textit{pruning patterns of individual columns} in the crossbars by the property \textbf{U}, we target the \textit{amount of sparsity across columns}. 
\textbf{D} dictates the base sparsity of the crossbar (i.e. the sparsity of the least sparse column), and the remaining columns will have greater or equal sparsity. 
However, the lower the sparsity (the number of zeroed-out weights) in the remaining columns, the better the accuracy, without any need for increased ADC precision. Hence, we demand all columns to get as close as possible to the least sparse columns, encouraging balanced sparsity across crossbar columns as can be seen from the two rightmost columns in Fig.~\ref{fig:sparsity}. We term this the \textbf{Balanced} property (\textbf{B}).

The desired \textbf{DUB} sparsity is achieved by our approach as the combination of a training scheme (preparing weights) and a post-training pruning scheme (zeroing out weights).
This is because, while sparsity with \textbf{D} property is enforced during pruning, \textbf{U} and \textbf{B} properties are attained as a result of DNN training.
Our method achieves this by regularizing the variance of $L_{0}$ norms of neighboring columns within the same crossbar. 
Our proposed implementation allows the regularization to be directly used in end-to-end gradient-based training.
This, in turn, allows the optimization process to decide different degrees of pruning tolerable by each crossbar dynamically and eliminates the need to pre-determine them before the training.
Furthermore, the proposed approach only requires information about crossbar sizes, allowing it to be applicable without the need for detailed information about the underlying crossbar-based architecture.

To summarize, this work makes the following contributions:
\begin{itemize}
    \item We identify three key properties that enable pruning in crossbars to reduce ADC energy without sacrificing accuracy.

    \item We propose a training method that minimizes the variance of $L_{0}$ norms of columns in the same crossbar to achieve both unstructured and balanced sparsity in order to amortize the accuracy drop.
    
    \item We prune each individual crossbar to ensure that sparsity translates to hardware efficiency by restricting sparsity levels to powers of $2$.
    
    \item We evaluate the accuracy-energy trade-offs of the proposed method on convolutional layers of VGG11 and ResNet18 models trained on CIFAR10 and ImageNet, respectively.
\end{itemize}
\section{Background and Related Works}
\subsection{Crossbar-based analog in-memory processing}\label{sec:background_xbars}
\begin{figure}[t!]
    \centering
    \includegraphics[width=0.98\linewidth]{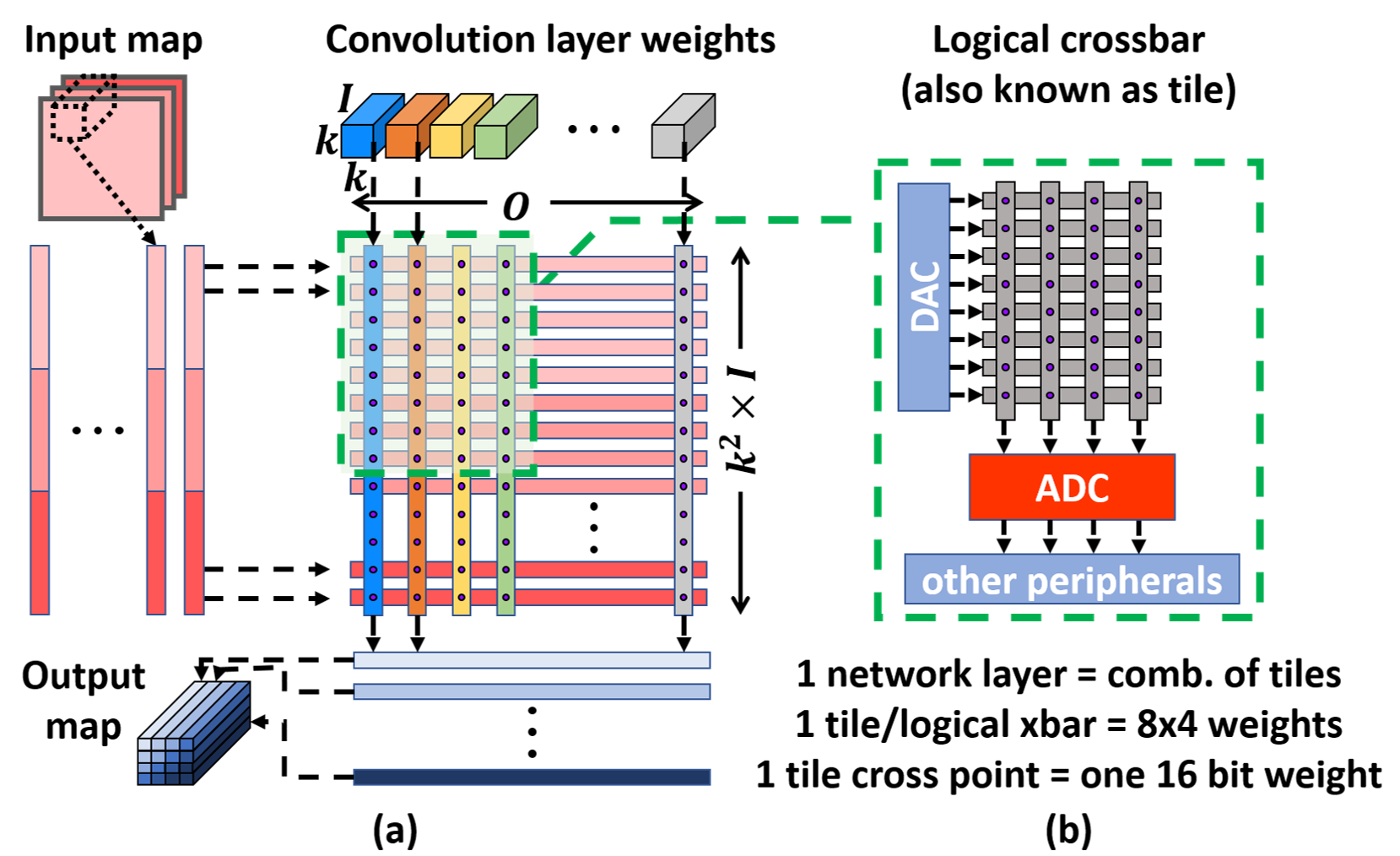}
    \caption{(a) Example illustrating how a convolutional layer of DNN is implemented as crossbar-based matrix-vector multiplication. (b) The logical crossbar (tile) structure that is usually used in machine-learning accelerators.
    }
    \label{fig:mapping}
\end{figure}

Crossbar-based analog in-memory architectures~\cite{shafiee2016isaac, chi2016prime, song2017pipelayer, ankit2019puma, ankit2020panther, wan2022compute, yang2019sparse, chen202115, ali202135, kim2023samba, ogbogu2023energy} improve DNN inference by implementing fast and efficient in-situ matrix-vector multiplication units, which can be used for both linear and convolutional layers.
Fig.~\ref{fig:mapping}(a) illustrates the general procedure of how the convolution operation of a single DNN layer is implemented as MVM using crossbar-based mapping. 
First, the weight parameters of the layer are stored (mapped) as device values at cross points of crossbars.
In the case of a convolutional layer, the weights are flattened into a $2$D matrix of width $O$ and height $(k^2\times I)$, where $O$ is the number of output filters, $I$ is the number of input channels, and $k$ represents both the width and height of the convolutional kernel.
Second, every $(k^2\times I)$ patch of an input map that needs to be convolved by each of the $O$ filters is flattened into the input vector and sequentially applied to the rows of the $2$D crossbars.
As a result, the in-place matrix-vector multiplication is performed by the application of input vectors to the rows of the stored $2$D weight matrix.
Each of the resulting $O$-dimensional points of the output map is then read from crossbar columns.

Various architectures $may$ have different weight mapping schemes of quantized bits of each weight onto \textit{physical crossbars}.
For example, while the ISAAC architecture~\cite{shafiee2016isaac} maps all bits of the same weight next to each other onto the same physical crossbar, the PUMA architecture~\cite{ankit2019puma} maps bits of the same weight onto separate physical crossbars.
Hence, operating with physical crossbars necessitates incorporating more details than just the crossbar size from the perspective of an algorithmic approach.

In order to design a general pruning method applicable to a wide range of architectures, in this work we think in terms of a more abstract structure than physical crossbars.
\textit{Logical crossbars}, also known as \textit{tiles} in various accelerator architectures, are the abstraction at which it can be assumed that each cross point represents one weight regardless of its required fixed precision.
Fig.~\ref{fig:mapping}(b) shows a general structure of a logical crossbar that can be thought to have a digital-to-analog converter (DAC), which converts inputs to voltage signals applied at the rows of crossbars, an analog-to-digital converter (ADC), which converts analog dot products produced by each column into digital partial sums, and other peripherals, such as shift-\&-add registers.
Based on the exact architecture design, each logical crossbar will be realized differently requiring a different number of physical crossbars and associated peripheral components.
In that case, the DAC and the ADC shown in Fig.~\ref{fig:mapping}(b) collectively represent a set of components rather than only a single peripheral component.
Moreover, it is justified to consider sparsity at the logical crossbar granularity, because removing weight from a logical crossbar will result in removing all cells (zeroing out devices) corresponding to the bits of that weight from all physical crossbars constituting this logical crossbar.
For example, if the tile gets at least $50\%$ sparsity on all of its columns, all the physical crossbars forming it will have at least $50\%$ sparsity on all of the columns too.

\subsection{Pruning methods}
\subsubsection{Neural network sparsity}
Pruning is an effective method to compress DNNs by means of zeroing out a large portion of network parameters (also known as \textit{inducing sparsity} into the network weights)~\cite{blalock2020what}. 
Based on the constraints put on the final sparsity patterns, pruning methods can be categorized into two categories: unstructured and structured.
Unstructured pruning~\cite{han2015deep, han2015learning, zhang2018systematic, rathi2018stdp} induces sparsity at the finest granularity by pruning the individual weights in neural networks.
By imposing a little constraint on weights, it hence offers large compression ratios at the cost of little accuracy drop.
However, unstructured methods are generally overshadowed by structured approaches due to the high irregularity of produced sparsity patterns and the difficulty of leveraging them.
Structured pruning~\cite{wen2016learning, li2016pruning, garg2019low, yang2020harmonious, aketi2020gradual} imposes stronger constraints demanding the removal of entire groups of weights, such as filters, channels, or entire layers in neural networks.
Despite it incurs high costs in the accuracy of DNNs when compared to unstructured sparsity~\cite{mao2017exploring}, such structured sparsity is easier to leverage in traditional hardware, such as modern GPUs or FPGA-based architectures.

While there are works taking advantage of such general neural network sparsity for crossbar-based analog in-memory architectures~\cite{yang2019sparse, chen202115, ali202135, kim2023samba, ogbogu2023energy}, the issue is that they require substantial changes either to the underlying architecture and/or to the data movement of input activations and weights. 
For example, the work in~\cite{kim2023samba} achieves computational speed up by enabling the rearrangement of weights according to the sparsity by making changes at the compiler level and utilizing extra crossbar arrays.
As a result, it is favorable to adapt algorithmic pruning approaches that demand less significant changes to the underlying architecture by inducing sparsity \textit{specifically for crossbar-based architectures}.

\subsubsection{Pruning for crossbars}
When underlying architecture utilizes crossbar-based mapping, it is possible to consider pruning at the fine granularity of a crossbar, crossbar rows, and crossbar columns.
Authors of \cite{liang2018crossbar} target reducing the number of required crossbars by either removal of entire crossbars or removal and rearrangement of crossbar columns pruned by $L_{0}$ norm constrained optimization.
Works in \cite{lin2019learning} and \cite{ankit2019trannsformer} target removing crossbars based on their utilization through column clustering and crossbar clustering, respectively.
Authors in \cite{chu2020pim} target reducing the number of crossbars by simultaneous pruning in crossbar row and crossbar column directions, requiring a change to the data path.
Authors in~\cite{yuan2021tinyadc} propose the pruning approach based on a similar type of sparsity as targeted in our work using the ADMM optimization approach, which, however, requires a predetermined number of non-zero weights in each crossbar column due to the rigidness of the optimization.
Similarly, authors in~\cite{xue2023hierarchical} also use ADMM optimization for weight pruning, but disregard balancing of the sparsity within crossbar columns in favor of heuristic-based bit pruning.

The advantages of our proposed method (and the resulting \textbf{DUB} sparsity) over the existing pruning techniques for crossbar-based architectures can be summarized by the following points.
\textbf{First}, our aim of reducing peripheral energy can be considered as an orthogonal problem to the problem of reducing the number of crossbars.
While the latter focuses on reducing the number of computations, our work focuses on reducing the cost of each computation in terms of ADC energy consumption, the major component of energy in crossbar-based architectures.
\textbf{Second}, our work leverages unstructured sparsity that is generally overlooked in favor of structured pruning.
This significantly lowers the constraint on weights, allowing for a much higher sparsity at a finer granularity than the structured sparsity considered before for crossbars.
Additionally, unstructured sparsity assumes no changes to the methodology of mapping weights to crossbars or data paths responsible for fetching input bits.
\textbf{Third}, the balancing of the sparsity within crossbar columns enabled by our method directly prepares DNN weights to gracefully amortize the accuracy for the final discretized and unstructured sparsity during training.
This is in contrast to other methods, where, without the balancing property, the optimization (or the fine-tuning) has to maximize the utilization of the remaining weights after the pruning is complete.
\textbf{Finally}, our proposed method is simpler in terms of implementation, such that the desired sparsity is achieved directly through the differentiable regularization viable in the end-to-end gradient-based training. 
Consequently, it does not require to set the degree of pruning before the training process. 
Instead, our approach allows each individual crossbar to have its own degree of pruning, which is determined by the overall importance of weights mapped to the crossbar as well as by the sparsity of the neighboring columns within the crossbar. 
As a result, a more flexible trade-off can be achieved by preserving high ADC precision levels on crossbars with the important weights, while aggressively pruning the others.

\section{DUB sparsity for ADC efficiency}
\subsection{\textbf{Discretized (D) sparsity}}
This work aims to reduce ADC energy in crossbar-based architectures \textit{while} maintaining DNN accuracy by inducing \textit{ADC-specific sparsity}, i.e. sparsity that translates into a reduction in the required ADC precisions. An ADC with $M$ bits precision has a resolution of $2^M$, resulting in $2^M$ discrete levels. In order to reduce ADC precision by $x$ bits, the resolution has to be limited to $2^{M-x}$ levels. This means that it is necessary to achieve a \textit{target sparsity} (per tile) of:
\begin{equation}
    TS(x~\text{bits}) = \frac{2^{M} - 2^{M-x}}{2^M} = (1-2^{-x})\times 100\%
\end{equation}

Since each tile has an associated (set of) ADCs, their precisions can be reduced by pruning \textit{each individual tile} so that they reach ADC-specific \textbf{discretized sparsity levels}. Moreover, within a tile, we assume that an ADC is shared between all columns to reduce area and energy at the cost of latency. Hence, its precision is dictated not by the overall tile sparsity, but only by the sparsity of \textbf{the least sparse column (LSC)}. Therefore, \textit{all columns} in the tile will have sparsity that at least equals to a target sparsity $TS$.
For instance, to get a reduction of 1 bit, the LSC needs $TS(1) = 50\%$ sparsity, and for a reduction of 2 bits, the LSC needs $TS(2) = 75\%$ sparsity. Sparsity between such levels, e.g. below $50\%$ and between $51-74\%$, does not impact the ADC precision. This is shown in the two leftmost columns of Fig.~\ref{fig:sparsity}. The LSC had a sparsity of $37.5\%$ which is pulled up to $50\%$ to allow reduction of the ADC precision by 1 bit. Since only the LSC affects the accuracy, the remaining columns can be used to regain accuracy according to the \textbf{U} and \textbf{B} properties.

\subsection{\textbf{Unstructured (U) sparsity}}
The \textbf{D} property does not pose any constraints on the pattern of pruned weights. A consequence of column-wise readout in the tile is that the locations of zeros are irrelevant, allowing a column to accommodate unstructured sparsity patterns that usually incur a lesser loss in accuracy than structured sparsity patterns. Different columns can have different sparsity patterns as long as the amount of sparsity in each column is greater than \textit{TS} of the tile.
This is illustrated in the 2 crossbar sparsity patterns shown in the middle of Fig.~\ref{fig:sparsity}.
While both crossbars have the same \textit{TS} of $TS(1)=50\%$, the crossbar with \textbf{U} property (on the right) enables fewer constraints by allowing free weight positioning.

\subsection{\textbf{Balanced (B) sparsity}}
Directly inducing tile-wise unstructured sparsity is not the best-suited trade-off for accuracy and hardware-efficiency. Since we do not know in advance which column will be the LSC after training, the intuitive approach is to induce as much sparsity in each column. However, since only the sparsity in the LSC dictates the ADC precision, excessively pruning the remaining columns does not result in additional hardware savings, but is likely to result in accuracy drops by limiting the number of useful DNN weights. Hence, in order to utilize the hardware well, without compromising accuracy, we want all columns to be pruned to the same degree as shown in the last two columns of Fig.~\ref{fig:sparsity}. We term this the \textbf{balanced property} since we wish to induce a similar degree of sparsity across intra-tile columns.

\begin{figure*}[ht]
    \centering
    \includegraphics[width=0.95\linewidth]{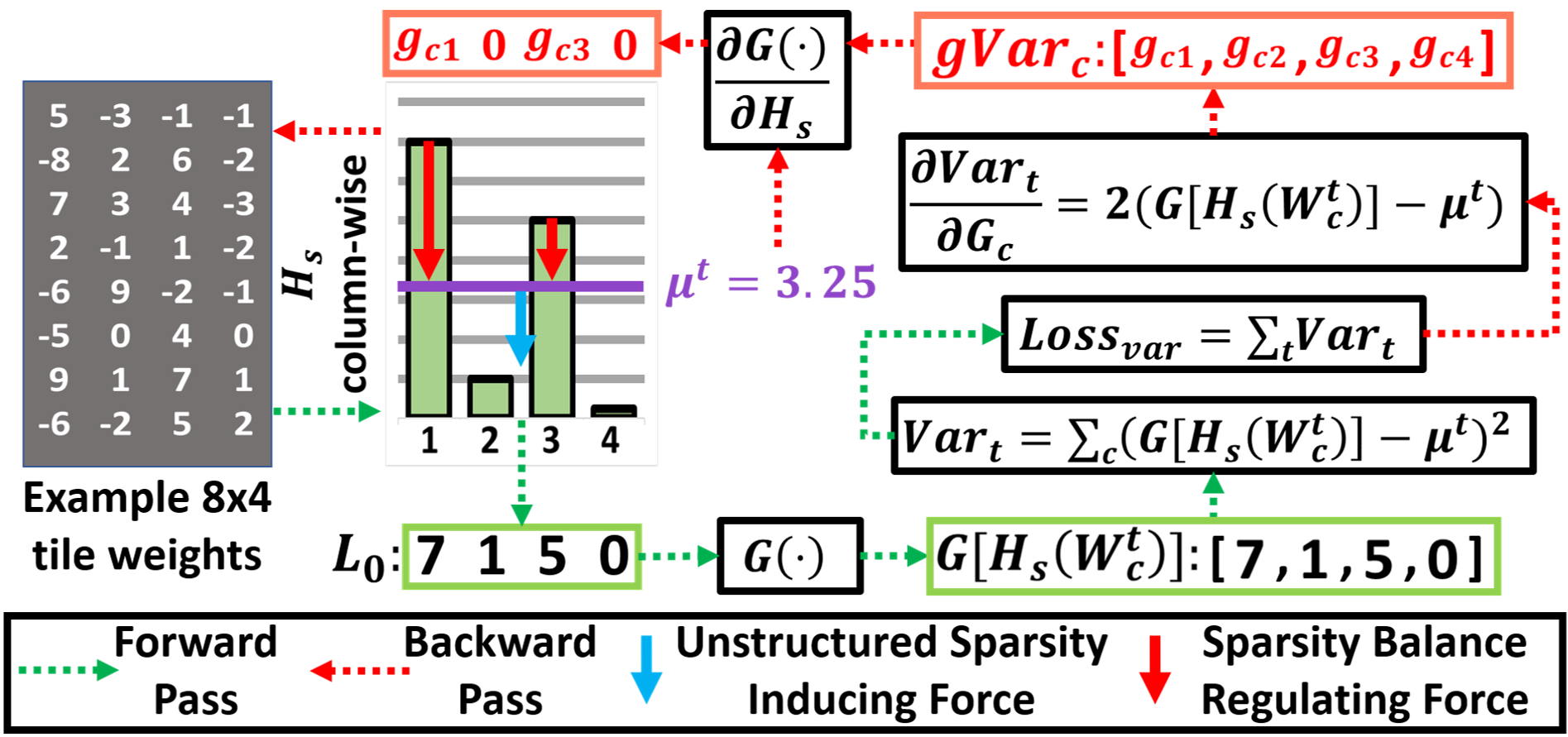}
    \caption{Example of training a $8\times4$ tile for \textbf{U and B} sparsity. Note how gradients due to variance $gVar_c = [g_{c1},g_{c2},g_{c3},g_{c4}]$ are zeroed out for columns 2 and 4 after passing gradient gate $\partial G(\cdot)/\partial H_s$ leaving balance regulating force \textbf{only} on columns 1 and 3, which have $L_{0}$ greater than $\mu^t$.
    }
    \label{fig:training}
\end{figure*}

\section{Methodology}
\subsection{Algorithm Overview}
The sparsity with \textbf{D} property can be directly enforced on the weights of a completely trained DNN purely based on any desired reduction in ADC precision.
However, the direct approach of removing weights after training limits its performance, because the recovery of accuracy solely depends on the fine-tuning of the weights remaining after pruning is complete. In other words, the training of DNN does not account for the removal of the weights, regardless of the desired granularity of sparsity.
Moreover, such an approach would have to either induce the same degree of sparsity in every crossbar~\cite{yuan2021tinyadc}, regardless of the susceptibility of the corresponding weights or rely on some heuristics~\cite{xue2023hierarchical} to deduce a different degree for each crossbar.

To that effect, our proposed method augments the training process of DNN weights, which aligns/prepares DNN weights such that the final resulting sparsity has the desired \textbf{U} and \textbf{B} properties along with the \textbf{D} property.
Moreover, the augmentation is implemented through the differentiable regularization approach, which allows the optimizer to automatically consider the possibility of inducing variable degrees of sparsity into each crossbar during end-to-end gradient-based training.
As a result, \textbf{DUB} sparsity is achieved by implementing the proposed approach to operate at two different phases: during training, and then during pruning.
We \textbf{first train} for \textbf{U and B} sparsity \textit{within each tile}, and in the second phase, we \textbf{then remove weights} to achieve the final \textbf{D, U, and B} sparsity \textit{per tile}.

Section~\ref{sec:training_for_u_and_b} describes the training process for \textbf{U and B} properties, using the newly proposed variance regularization of $L_{0}$ norms of neighboring columns and gradient gating. Then, Section~\ref{sec:pruning} describes the proposed pruning process of a fully trained DNN to induce \textbf{D} sparsity, while preserving \textbf{U and B} properties, based on the susceptibility of weights within each tile to a pruning threshold and discretized sparsity levels attuned to the reduction of ADC precision.

\subsection{Training for intra-tile \textbf{U and B} sparsity}\label{sec:training_for_u_and_b}
\subsubsection{Achieving U and B sparsity}
Training for \textbf{U} and \textbf{B} sparsity across the tile columns is implemented by exerting two forces on a tile as shown in Fig.~\ref{fig:training}: First, \textit{an overall unstructured sparsity inducing force} (blue arrow in Fig.~\ref{fig:training}) and second, \textit{a per-tile sparsity balance regulating force} (red solid arrows in Fig.~\ref{fig:training}). We explain their roles and implementation as follows.

\textit{The overall unstructured sparsity-inducing force} has the purpose of driving the mean sparsity of a tile. It was chosen to be the standard weight decay through $L_{2}$ regularization applied on all weights. (We chose $L_{2}$ rather than $L_{1}$ regularization for this purpose as the former was shown to have a better effect after retraining~\cite{han2015learning}.) 
However, since this regularization is unaware of the crossbar structure, it neither targets the LSC of each tile nor enforces the balanced property.
These require varying levels of force to be applied to different columns based on their current sparsity levels. Hence, \textit{the balancing force} operates on each tile individually to impose the two desired goals. The first goal is to induce more sparsity on the columns that are more dense in anticipation of one of them becoming the LSC after training. The second goal is to achieve a balance in sparsity across the columns by \textbf{not} inducing sparsity on less dense columns.

In order to accommodate both these goals, we evaluate each column individually with respect to the overall mean density of the columns in the tile. The density is measured by the $L_{0}$ norm (i.e. the number of non-zero elements), and we denote the mean value of the $L_{0}$ norms of all columns in the tile $t$ by $\mu^t$. We then enforce \textbf{more sparsity} on the columns that have an $L_{0}$ norm \textbf{greater than $\mu^t$}. For example, in Fig.~\ref{fig:training}, $L_{0}$ norms of columns 1 and 3 are higher than $\mu^t$, and therefore they receive additional regularization force over the standard $L_{2}$, in accordance to the first goal. Second, it is desirable to attain \textbf{B} sparsity to avoid underutilized tiles. This is achieved by \textbf{not inducing additional sparsity} on the columns that have $L_{0}$ norm \textbf{lower than  $\mu^t$}. For example, in Fig.~\ref{fig:training}, $L_{0}$ norms of columns 2 and 4 are lower than $\mu^t$ and therefore receive only the standard $L_{2}$ regularization, achieving the second goal. The mean updates with each change, and eventually will settle at a value that encourages balanced sparsity across columns.

In our algorithm, we implement the balancing force by minimizing the variance of the $L_{0}$ norms of the columns in a tile with the special condition, which we term ``gradient gating''. Specifically, ``gradient gating'' passes the gradients only to weights of those columns that have $L_{0}$ norm higher than $\mu^t$, while zeroing out gradients on the rest of the tile columns. 
In the context of Fig.~\ref{fig:training}, the gradient gating mechanism zeroes out the gradients of columns 2 and 4 (due to balancing force, but not $L_{2}$ force) during the backward pass. However, these columns still indirectly affect the intensity of regularization applied to the other columns (1 and 3) by virtue of effectively driving the value of $\mu^t$ lower.

\subsubsection{Challenge of estimating training-time sparsity}
Determining the number of zero weights in each column presents a challenge during training. We use the $L_{0}$ norm, which tells us the count of non-zero weights during the forward pass. However, incorporating it into gradient-based learning algorithms is a challenge since it is not differentiable. Hence, we use the recently presented Hoyer-Square~\cite{yang2019deephoyer} metric as the differentiable approximate measure of the $L_{0}$ norm. It is calculated as the ratio of the square of the $L_{1}$ norm and the $L_{2}$ norm and approximates the $L_{0}$ norm while being differentiable. Interested readers are encouraged to refer to the original work~\cite{yang2019deephoyer} for full details. 
If $\textbf{w}$ represents the weight vector of a tile column, then the Hoyer-Square applied to the column is calculated as: 
\begin{equation}\label{eq:HoyerSquare}
    L_{0}(\textbf{w}) \approx H_s(\textbf{w}) = \frac{(\sum_i |\textbf{w}_i|)^2}{\sum_i \textbf{w}_i^2}
\end{equation}

\subsubsection{Gradient gating and loss functions}
We implement ``gradient gating'', denoted as $G(\cdot)$, as a custom PyTorch function and apply it on top of the Hoyer-Square measure before it is used to compute the variance. This is shown in Fig.~\ref{fig:training}. 
We represent weights of the column $c$ in tile $t$ as $\textbf{W}^t_c$.
During the forward pass, represented by green dashed arrows, $G(\cdot)$ just passes through the value of $H_{s}(\textbf{W}^t_c)$. 
We then compute the variance of the gated $H_{s}(\textbf{W}^t_c)$ values of all $\mathcal{C}$ columns mapped together to tile $t$. 
During the backward pass, represented by the red dashed arrows, $G(\cdot)$ conditionally zeroes out (or passes-through) incoming gradients from the variance, $gVar_{c}$, based on whether the $H_{s}$ measure of the column $c$ was lower (or higher) than the mean of $H_{s}$ measures, denoted by $\mu^t$ and depicted by the purple line. Gradient gating is formulated as:
\begin{equation}\label{eq:GatedHoyerSquare}
\frac{\partial G(H_s(\textbf{W}^t_c))}{\partial H_s} = \left\{
    \begin{array}{ll}
        gVar_{c} & \mbox{if } H_s(\textbf{W}^t_c) > \mu^t \\
        0 & \mbox{otherwise}
    \end{array}
\right.
\end{equation}
where $\mu^t = \frac{1}{C}(\sum^{\mathcal{C}}_c H_s(\textbf{W}^t_c))$ is the mean of $H_{s}$ measures of all $\mathcal{C}$ columns of some tile $t$.
These gradients are then passed back to change the individual weights of the corresponding columns since the $H_{s}$ measure is differentiable. 

The training loss function to achieve for intra-tile \textbf{U} and \textbf{B} sparsity is formulated as shown:
\begin{equation}\label{eq:training}
\begin{split}
    loss & =  loss_{cls} 
             + \lambda_{mean} \cdot \sum^{N}_n \textbf{W}^2_n \\
            & + \lambda_{var} \cdot \sum^T_{t} \underbrace{\sum^{\mathcal{C}}_{c} \left(G \Big[ H_s(\textbf{W}^t_c) \Big] - \mu^t \right)^2}_{tile-wise~variance~of~L_{0}~norms}\\
\end{split}            
\end{equation}
where $loss_{cls}$ denotes a classification loss (cross-entropy in our experiments) and $\textbf{W}$ represents the network weights with the total number of weights $=N$. $T$ denotes the total number of required tiles, each with $\mathcal{C}$ number of tile columns.
Hyperparameters $\lambda_{mean}$ and $\lambda_{var}$ define the trade-off between classification loss, network-wide unstructured sparsity, and the balance of unstructured sparsity within each tile.

\subsection{Pruning for per-tile \textbf{D} sparsity}\label{sec:pruning}
In the previous subsection, we describe how we train in order to encourage pruning attuned to the crossbar structure. After the training is complete, we perform pruning on the per-tile basis with considerations of both the layer-wise pruning threshold and sparsity levels suitable to ADC precision levels. 
Generally, in most pruning algorithms, we empirically find layer-wise pruning thresholds, and weights below this threshold are zeroed out. 
In the crossbar structure, the layer-wise thresholds translate to certain sparsity levels per column in each tile, but we enforce the discretized \textbf{D} property prior to actually zeroing out the weights.
In order to do that, we \textit{evaluate each tile individually} to identify the LSC \textit{as per the layer-wise threshold}. 
This may not adhere to the ADC precision levels that dictate the acceptable discretized sparsity. 
Hence, we round this value to the closest ADC-attuned discretized level. 
All the columns in the tile are then pruned to \textit{the same discretized sparsity level} corresponding to the determined precision from the rounded discretized sparsity of LSC.
Note that, because all columns would have a higher sparsity than the LSC (by its definition) if pruned according to a generic layer-wise threshold, pruning them instead to the same sparsity level of the LSC preserves more number of parameters. 
This allows us to regain the accuracy by preserving property \textbf{B} encouraged by the proposed training.

As an example, consider a tile with dimensions $64\times64$. Based on the threshold for the layer, certain weights are identified per column per tile for removal. Let's assume that for a particular tile in the layer, the LSC is determined to have $45/64 \approx 70\%$ zeroed weights. The closest ADC-attuned discrete sparsity level is $75\%$, which corresponds to reducing the ADC precision of that tile by $2$ bits. 
All other columns will have identified zeroed weight percentages $> 70\%$. 
Let's say the second least sparse column was identified to have $55/64$ or ($\approx 86\%$) weights to be pruned.
This means that according to the generic layer-wise pruning scheme, this column will retain only $64-55=9$ non-zero weights.
However, according to our per-tile pruning scheme, in \textbf{all columns} of this tile $75\%$ or $48$ of smallest magnitude weights will be pruned.
As a result, there are $25\%$ or $16$ weights \textbf{in each of the 64 columns} that can be used to help recover accuracy during fine-tuning.
This includes the second least sparse column, which would otherwise have only $9$ non-zero weights if pruned according to the generic scheme.

\begin{table*}[ht!]
\centering
\caption{Accuracy, Normalized ADC energy, and Pruning Ratio achieved by different training and pruning approaches for VGG11 and ResNet18 trained on CIFAR10 and ImageNet datasets, respectively, with tile size of $64\times64$ weights}
\label{tabel:accuracyvsenergy}
\resizebox{\linewidth}{!}
{%
\begin{tabular}{|c|l|c|c|c|c|c|c|} 
\hline
\multicolumn{2}{|l|}{Training and pruning method} & Accuracy & \begin{tabular}[c]{@{}c@{}}Accuracy Difference\\w.r.t. Baseline\end{tabular} & \begin{tabular}[c]{@{}c@{}}Normalized ADC \\Energy\end{tabular} & \begin{tabular}[c]{@{}c@{}}ADC Energy \\Savings\end{tabular} & Allowed Pruning Ratio & Final Pruning Ratio \\ 
\hline
\multicolumn{8}{|c|}{\textit{VGG11 model trained on CIFAR10 dataset}} \\ 
\hline
\multicolumn{2}{|l|}{Baseline} & 90.87 & - & 1.00 & - & - & - \\ 
\hline
\multicolumn{2}{|l|}{Unstructured pruning~\cite{han2015learning}} & 90.26 & -0.61 & 0.384 & 2.60x & 95.65\% & 95.65\% \\ 
\hline
\multicolumn{2}{|l|}{\textbf{Our method}} & \textbf{90.30} & \textbf{-0.57} & \textbf{0.250} & \textbf{4.00x} & \textbf{95.65\%} & \textbf{90.19\%} \\ 
\hline
\multirow{6}{*}{\rotatebox[origin=c]{90}{Structured methods~~~~~}} & \begin{tabular}[c]{@{}l@{}}Crossbar column pruning\\(same sparsity structure as in~\cite{liang2018crossbar, lin2019learning, chu2020pim})\end{tabular} & 90.28 & -0.59 & 0.412 & 2.43x & 95.65\% & 95.65\% \\ 
\cline{2-8}
 & \begin{tabular}[c]{@{}l@{}}Crossbar row pruning\\(same sparsity structure as in~\cite{chu2020pim})\end{tabular} & 90.71 & -0.16 & 0.323 & 3.10x & 95.65\% & 95.65\% \\ 
\cline{2-8}
 & \begin{tabular}[c]{@{}l@{}}Crossbar tile pruning\\(same sparsity structure as in~\cite{liang2018crossbar, ankit2019trannsformer, chu2020pim})\end{tabular} & 90.63 & -0.24 & 0.284 & 3.53x & 95.65\% & 95.65\% \\ 
\cline{2-8}
 & \begin{tabular}[c]{@{}l@{}}Training for crossbar column sparsity but with\\ \textbf{our per-tile pruning for D and B sparsity}\end{tabular} & 90.79 & -0.08 (\textbf{B}) & 0.346 & 2.89x (\textbf{D}) & 95.65\% & 86.88\% \\ 
\cline{2-8}
 & \begin{tabular}[c]{@{}l@{}}Training for crossbar row sparsity but with\\ \textbf{our per-tile pruning for D and B sparsity}\end{tabular} & 91.09 & +0.22 (\textbf{B}) & 0.282 & 3.55x (\textbf{D}) & 95.65\% & 88.36\% \\ 
\cline{2-8}
 & \begin{tabular}[c]{@{}l@{}}Training for crossbar tile sparsity but with\\\textbf{our per-tile pruning for D and B sparsity}\end{tabular} & 91.06 & +0.19 (\textbf{B}) & 0.254 & 3.94x (\textbf{D}) & 95.65\% & 88.56\% \\ 
\hline
\multicolumn{8}{|c|}{\textit{ResNet18 model trained on ImageNet2012 dataset}} \\ 
\hline
\multicolumn{2}{|l|}{Baseline} & 69.76 & - & 1.00 & - & - & - \\ 
\hline
\multicolumn{2}{|l|}{Unstructured pruning~\cite{han2015learning}} & 69.16 & -0.60 & 0.910 & 1.10x & 72.74\% & 72.74\% \\ 
\hline
\multicolumn{2}{|l|}{\textbf{Our method}} & \textbf{68.90} & \textbf{-0.86} & \textbf{0.787} & \textbf{1.27x} & \textbf{72.74\%} & \textbf{56.85\%} \\ 
\hline
\multicolumn{2}{|l|}{\begin{tabular}[c]{@{}l@{}}Crossbar tile pruning\\(same sparsity structure as in~\cite{liang2018crossbar, ankit2019trannsformer, chu2020pim})\end{tabular}} & 68.54 & -1.22 & 0.930 & 1.07x & 72.74\% & 72.74\% \\
\hline
\multicolumn{2}{|l|}{\begin{tabular}[c]{@{}l@{}}Training for crossbar tile sparsity but with\\ \textbf{our per-tile pruning for D and B sparsity}\end{tabular}} & 68.92 & -0.84 (\textbf{B}) & 0.829 & 1.21x (\textbf{D}) & 72.74\% & 50.50\% \\
\hline
\end{tabular}
}
\vspace{-3.0mm}
\end{table*}

\section{Results}
\subsection{Evaluation criteria}
In the previous section we described algorithmic optimization that targets ADC energy requirements of crossbar-based architecture.
Our method achieves this by pruning weights based \textbf{only} on the information about \textbf{crossbar sizes}.
During evaluations, we consider square tiles of shape $n\times n$ for values of $n\in\{32, 64, 128, 256\}$.
Because the primary focus is to improve ADC energy, the effectiveness of our method was evaluated by estimating the normalized ADC energy savings instead of the absolute total energy savings.
This allows more general evaluation without the need of the knowledge about specifics of the target crossbar architecture and the fraction of energy ADCs consume as a part of the processing core within that architecture.
The normalized ADC energy is computed as a linear function of precision~\cite{murmann2020adc, ankit2019puma}, and can be formulated as:
\begin{equation}\label{eq:adc_energy}
    Norm.~E_{ADC} = \frac{1}{N}\sum^\mathcal{B}_{b=1} N_{b}\cdot \frac{b}{\mathcal{B}}
\end{equation}
where $N$ is the total number of tiles required by the network, $N_{b}$ is the number of tiles requiring ADC with $b$ bit precision, and $\mathcal{B}$ is the full ADC precision for the corresponding crossbar size.
Note that we consider a zero cost in ADC energy for tiles that are entirely pruned (have all weights pruned) and tiles that have only one non-zero weight on all of their columns.

To evaluate the effectiveness of the proposed method, we conduct two sets of experiments both of which are applied to convolutional layers of VGG11~\cite{simonyan2014very} and ResNet18~\cite{he2016deep} networks trained on CIFAR10~\cite{krizhevsky2009learning} and ImageNet2012~\cite{russakovsky2015imagenet} datasets, respectively.
Section~\ref{sec:first_experiment} presents the first experiment set, which analyzes the benefits of our method and the resulting \textbf{DUB} sparsity on ADC savings.
The proposed method is compared to the pruning of individual weights and the pruning of crossbar structures, such as tiles, tile columns, and tile rows.
Section~\ref{sec:second_experiment} presents the second experiment set, which explores how the benefits offered by our method can be combined with the benefits offered by structured pruning of entire crossbar tiles.

\subsection{Effectiveness of \textbf{DUB} sparsity on ADC savings}
\label{sec:first_experiment}
\subsubsection{Accuracy-energy trade-off}
We trained networks using various methods to induce sparsity at different scales and compared them based on normalized ADC savings and accuracy.
Our proposed method was used to induce \textbf{DUB} sparsity that is expected to be the most suitable for ADC energy savings. 
The pruning method presented in~\cite{han2015learning} was used to induce unstructured sparsity by removing individual weights, which is the finest possible sparsity regardless of a target hardware architecture.
The pruning method in~\cite{wen2016learning, yuan2006model} (\textit{Group Lasso}) was adapted to induce structured sparsity by removing crossbar structures. 
We considered the crossbar structures which were previously used for crossbar compression. 
Specifically, structured sparsity was induced by grouping and pruning crossbar columns as in~\cite{liang2018crossbar, lin2019learning, chu2020pim}, crossbar rows as in~\cite{chu2020pim}, or crossbar tiles as in~\cite{liang2018crossbar, ankit2019trannsformer, chu2020pim}.
To have a fair comparison, we allowed the same number of parameters to be pruned (\textbf{allowed pruning ratio}) as that achieved by the unstructured pruning and then selected hyperparameters so that accuracies were approximately similar with less than $1\%$ drop with respect to the baseline model.

Table~\ref{tabel:accuracyvsenergy} shows the accuracy-energy trade-off achieved by different training and pruning methods with respect to unpruned baseline networks for the tile size of $64\times64$ weights.
In the case of VGG11, our method achieves $4\times$ improvement in ADC energy in contrast to $2.6\times$ achieved with the unstructured pruning method with respect to the fully dense baseline model. 
This result illustrates that directly inducing unstructured sparsity is not enough. Instead, incorporating the knowledge in the form of crossbar size and discretized sparsity levels allows harnessing the unstructured sparsity with many favorable benefits for the reduction in ADC precision.
Both crossbar row and crossbar tile pruning methods achieve lesser improvements of $3.1\times$ and $3.53\times$, respectively, in ADC energy even under aggressive pruning ratio of $95.65\%$ weights.
Since both methods uniformly remove the weights from all columns in the crossbar, hence possibly achieving sparsity aligned with discretized levels, their improvement is better than that of the unstructured method. However, the rigid sparsity patterns only allow this much improvement within the budget of the allowed accuracy drop and the allowed pruning ratio. This signifies the importance of allowing unstructured and balanced sparsity, which enables more freedom in the distribution of remaining non-zero weights.
Finally, crossbar column pruning achieves the lowest improvement of $2.43\times$ in ADC energy. 
This is because such pruning only creates tiles that have some of the crossbars removed, while other columns remain completely dense. Consequently, except for some tiles that have all columns removed, it ensures neither discretized nor balanced sparsity, requiring high ADC precision.
Cumulatively, the presented results indicate that inducing \textbf{DUB} sparsity is indeed more beneficial for ADC energy savings than other types of sparsity.

\subsubsection{Effects of per-tile pruning with \textbf{D and B} sparsity}
Desired \textbf{DUB} sparsity is achieved by (1) training model weights so that they are aligned for \textbf{U and B} sparsity and (2) per-tile pruning of weights to discretized levels that enables \textbf{D, U, and B} sparsity together. 
Hence, we additionally analyzed the individual effect of our per-tile pruning and the resulting \textbf{D and B} sparsity when combined with the training methodology used for structured pruning methods. 
Specifically, instead of standard weight magnitude-based layer-wise pruning, we applied our per-tile pruning (section~\ref{sec:pruning}) to the networks trained by grouping~\cite{wen2016learning} together crossbar structures (i.e. crossbar columns, rows, or tiles).
The layer-wise thresholds that identify the number of weights allowed to be pruned (\textbf{allowed pruning ratio}) during per-tile pruning were still chosen to match the layer-wise compression ratios achieved by the unstructured pruning.
However, since our method additionally enforces sparsity attuned to ADC precision, we choose the final threshold for each individual tile accordingly. 
As explained in section~\ref{sec:pruning}, this results in less aggressive pruning (\textbf{final pruning ratio}) than standard weight magnitude-based layer-wise pruning, while being more or equally hardware-efficient.
Furthermore, because our per-tile pruning removes the same number of weights in the columns of the same crossbar, the balanced \textbf{B} property still holds, allowing extra weights to be re-utilized during the fine-tuning of the remaining weights.

Results of applying our per-tile pruning to the networks trained for the removal of crossbar structures are shown in \textbf{bold} in Table~\ref{tabel:accuracyvsenergy}.
For example, in the case of the crossbar tile sparsity, per-tile pruning boosted ADC energy savings from $3.53\times$ to $3.94\times$ while simultaneously improving accuracy from $90.63\%$ to $91.06\%$ (attaining higher accuracy than the baseline).
A similar pattern is observed in all three cases of model weights being trained to prune crossbar structures.
These results support our arguments for \textbf{D and B} sparsity.
By considering each individual tile and forcing them to have discretized (\textbf{D}) sparsity levels attuned to ADC precisions, per-tile pruning improves ADC energy savings.
By pruning all columns within tiles to the same balanced (\textbf{B}) sparsity, saved extra weights (note lower final pruning ratios) are re-utilized to better recover accuracy during fine-tuning.
However, note that, without the proposed training for unstructured sparsity with balanced (\textbf{U and B}) property, the proposed per-tile pruning has a limit on the amount of achievable ADC savings. This signifies the importance of aligning DNN weights through the training for the desired sparsity.

In the case of ResNet18, our method achieves an improvement in ADC energy of $1.27\times$ compared to $1.1\times$ of unstructured pruning and $1.21\times$ of the combination of training for crossbar tile removal with our per-tile pruning. 
Energy improvements are lower compared to VGG11, presumably owing to the general difficulty of pruning ResNet models. 
In the case of both network models, our training and pruning method achieves higher ADC energy savings and similar or lower accuracy drop than unstructured pruning unaware of target hardware and structured methods inducing coarser sparsity at crossbar structures.

\subsubsection{Sparsity within tiles}\label{sec:sparsity_within_tiles}
\begin{figure}[t!]
    \centering
    \subfloat[\label{fig:tiledist_cifar10}]{%
         \includegraphics[width=0.95\linewidth]{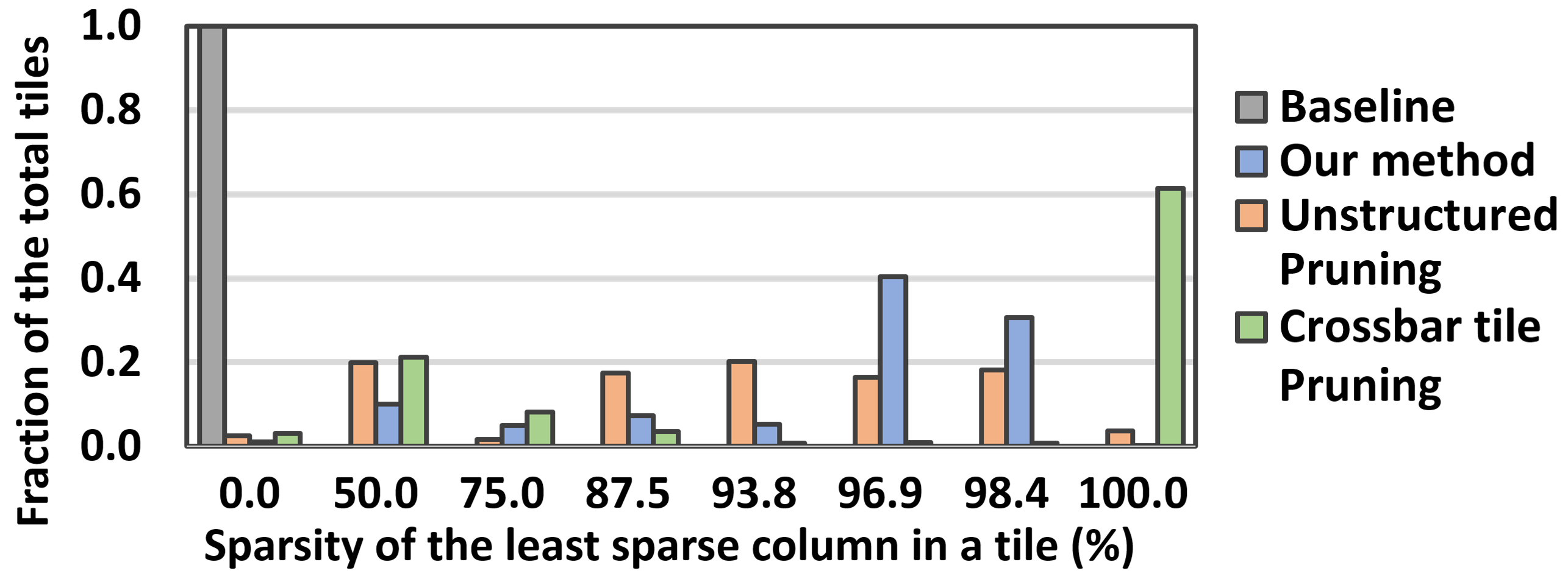}}
    \vspace{0.0mm}
    \subfloat[\label{fig:tiledist_imagenet}]{%
         \includegraphics[width=0.95\linewidth]{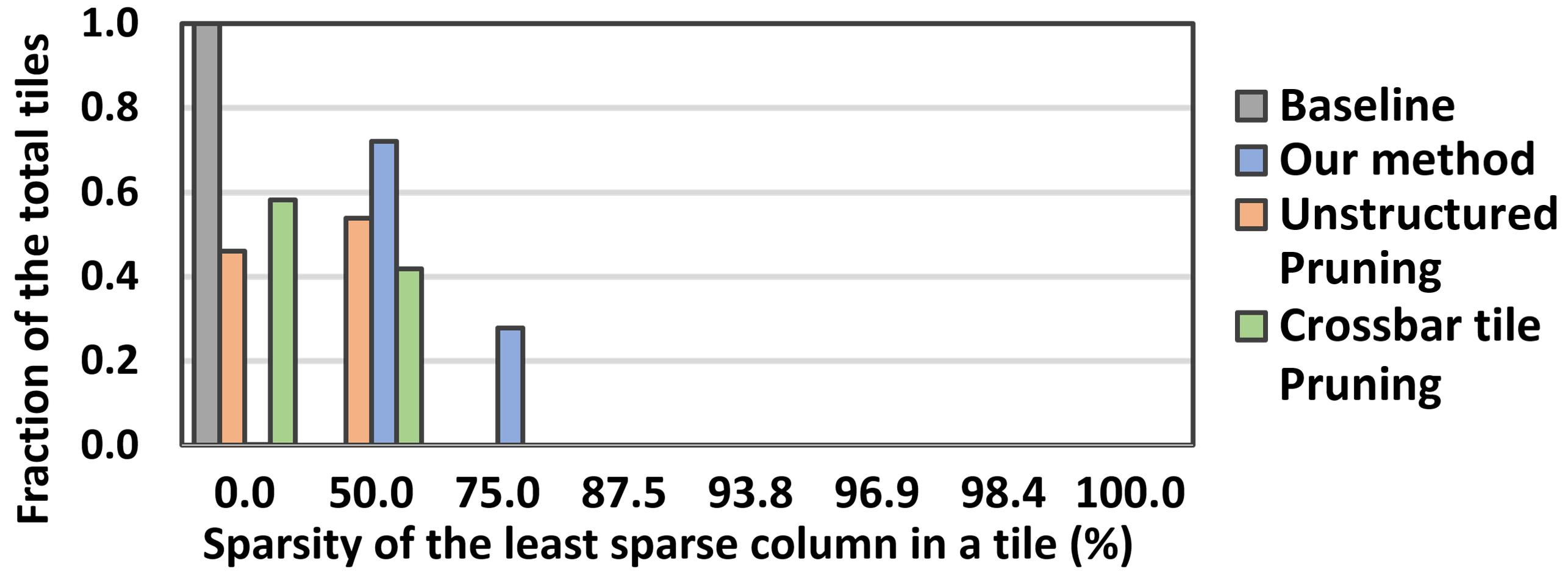}}
    \caption{Tile distribution based on their the least sparse columns obtained by different pruning methods applied to (a) VGG11 network trained on CIFAR10 and (b) ResNet18 network trained on ImageNet.}
    \vspace{-3.0mm}
    \label{fig:tiledist}
\end{figure}

In order to obtain more details on how \textbf{DUB} sparsity results in higher ADC savings and better accuracy, we analyzed the sparsity of tiles after training and pruning is complete.
We looked into the distribution of tiles based on their achieved target sparsity (\textit{TS}), which is determined by the corresponding least sparse column (LSC).
Figs.~\ref{fig:tiledist}(a)-(b) illustrate the distribution of tiles based on their LSC for VGG11 and ResNet18, respectively.
When compared with unstructured pruning, in both cases, our method is more effective at driving a larger fraction of tiles to have a higher sparsity on the LSC while still allowing irregular pruning patterns.

When compared with structured pruning, we should note that our method does not explicitly require entire tiles to be completely removed (i.e. $100\%$ sparsity).
Nevertheless, in the case of VGG11, about $30\%$ of all tiles can skip ADC computation ($\geq98.44\%$ sparsity) and about $40\%$ require only one-bit ADCs ($96.88\%$ sparsity).
In contrast, structured pruning aggressively removes more than $60\%$ of tiles and hence requires a larger number of highly dense tiles ($<75\%$ sparsity) with high ADC costs to compensate for the accuracy drop.
In the case of ResNet18, our method is able to almost entirely remove the requirement for full precision ADCs and to achieve $75\%$ target sparsity on one-third of all tiles enabling a reduction in the precision of their ADCs by $2$ bits.

The reason is that our proposed regularization enables different tiles to determine the optimal sparsity based on a ``majority'' decision: if the majority of the columns in a tile tend towards higher sparsity, the more persistent columns (the candidates to become LSC) are pushed along with the majority by the help of additional regularization from the balancing force (Section~\ref{sec:training_for_u_and_b}).
This is in contrast to both unstructured pruning and structured tile pruning. 
Purely unstructured pruning induces completely free irregular sparsity, which evenly spreads over different sparsity levels, due to the absence of any concern for the underlying crossbar-based segregation. 
On the other hand, structured pruning induces very rigid structured sparsity-based completely on the ``all or nothing'' principle, where it attempts to remove as many tiles as it can, but otherwise gives up on any tile with partial sparsity.
As a result, it can be deduced that our method achieves better ADC energy savings by naturally finding the balance between entirely pruning the tiles and sparsifying the rest of the tiles to reduced ADC precisions.

\subsubsection{Different tile sizes}
\begin{figure}[t!]
    \centering
    \subfloat[\label{fig:scalability_sdub}]{%
         \includegraphics[width=0.99\linewidth]{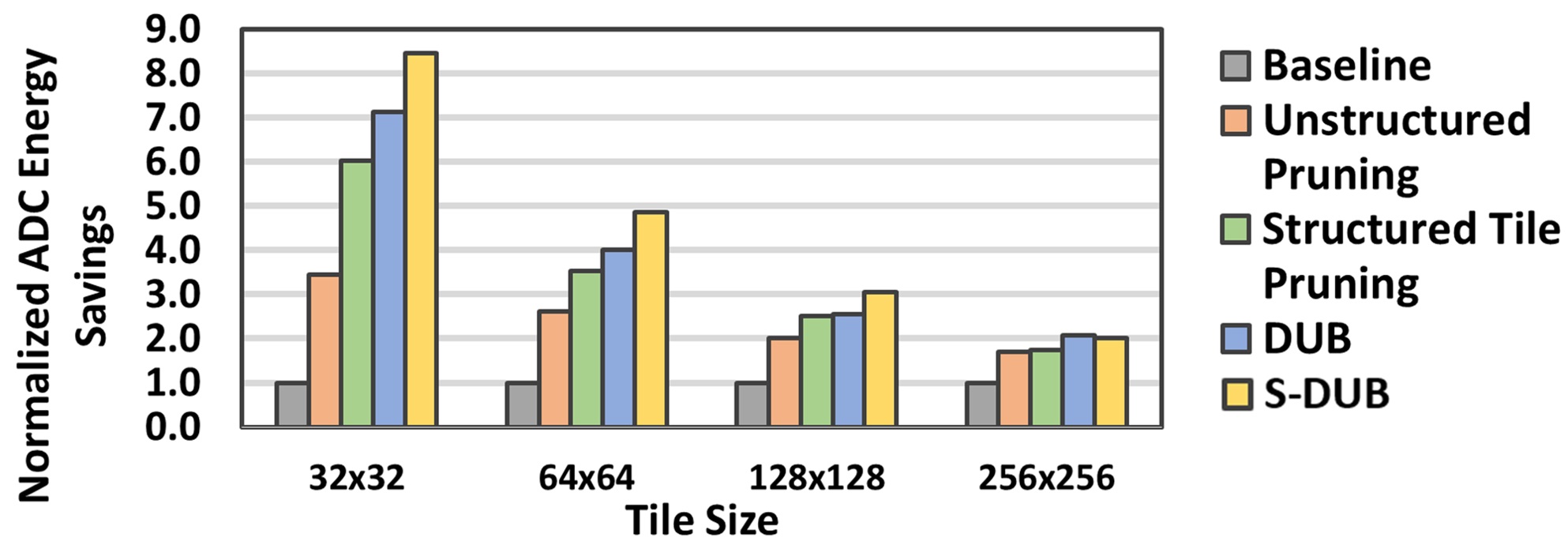}}
    \vspace{0.0mm}
    \subfloat[\label{fig:crossbar_removal_sdub}]{%
         \includegraphics[width=0.99\linewidth]{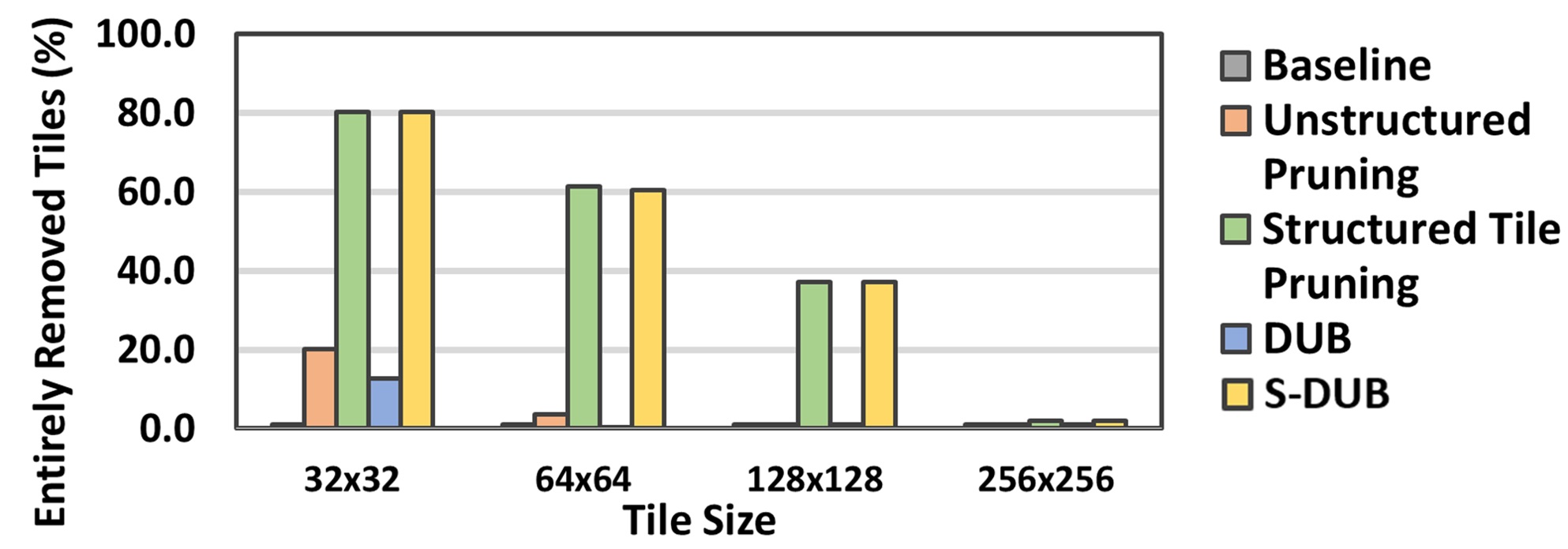}}
    \caption{(a) Normalized ADC savings and (b) fraction of entirely removed tiles for different tile sizes by different pruning methods on a VGG11 network trained on the CIFAR10 dataset.}
    \vspace{-3.0mm}
    \label{fig:scalability_and_crossbar_removal}
\end{figure}

We also analyzed the effectiveness of our method on different tile sizes by training VGG11 on CIFAR10 with $n\times n$ square tiles for various values of $n$.
Similar to the experiments above, we restricted all methods to have the same \textbf{allowed pruning ratios} and selected hyperparameters so that the accuracy of every method is within $1\%$ of the baseline accuracy.
Fig.~\ref{fig:scalability_and_crossbar_removal}(a) shows the ADC energy savings for different values of $n = 32,64,128,256$.
The best improvements in ADC energy can be observed for the tile size of $32\times32$ weights, where \textbf{DUB} sparsity-based ADC savings is $7.13\times$ the baseline.
This is in contrast to $6.0\times$ savings with tile pruning and $3.44\times$ savings with purely unstructured pruning.
However, the effectiveness of all methods reduces as the tile size increases. 
This can be explained by empirical observations~\cite{mao2017exploring, chu2020pim} that the difficulty of pruning increases with the increase in the crossbar array size since the increasing number of weights is tied in a crossbar structure.
Notably, the savings achieved by \textbf{DUB} sparsity is always greater than that of the other methods.

\subsection{Combination of our method with structured tile pruning}
\label{sec:second_experiment}
As pointed out earlier, \textbf{DUB} sparsity does not explicitly target the removal of entire tiles (i.e. $100\%$ sparsity), leading to balanced sparsity and high re-utilization of weights.
However, it is important to consider scenarios where a target crossbar-based architecture can be designed to skip computations of entirely pruned crossbar tiles.
Hence, we analyzed the performance of our method to prune entire tiles.
Fig.~\ref{fig:scalability_and_crossbar_removal}(b) compares the percentage of tiles removed by our \textbf{DUB} pruning technique, unstructured pruning, and structured tile pruning.
As expected, structured tile pruning performs better than our method in terms of removing entire tiles, due to the difference in their behavior explained in Section~\ref{sec:sparsity_within_tiles}.
This is especially evident in cases of small tiles, where the structured approach is capable of removing up to $80\%$ of $32\times32$ sized tiles and $61\%$ of $64\times64$ sized tiles.

To address the scenarios where a target architecture can skip computations of entire crossbar tiles, we propose to apply our method \textbf{after} eliminating entire tiles using structured tile pruning.
We refer to this approach as \textbf{S-DUB}.
On one hand, structured pruning offers the removal of a larger number of tiles, but at the cost of leaving a large fraction of tiles requiring high precision ADCs.
On the other hand, our method is effective in reducing ADC precision by skewing tile distribution towards higher sparsity levels as shown in Figs.~\ref{fig:tiledist}(a)-(b).
Hence, by combining two methods it is possible to achieve both the benefits of ADC savings and crossbar compression.

To verify the effectiveness of \textbf{S-DUB}, we performed structured tile pruning as in the previous experiments and then applied our method (both training and per-tile pruning) to the remaining tiles.
Fig.~\ref{fig:scalability_and_crossbar_removal}(a) and Fig.~\ref{fig:scalability_and_crossbar_removal}(b) show ADC savings and the percentage of tiles removed, respectively, using \textbf{S-DUB} in comparison with other methods.
Application of \textbf{S-DUB} adds considerable benefits in terms of ADC energy savings.
In the case of $32\times32$ tiles, improvements in ADC were boosted from $6.0\times$ (due to structured tile pruning) and $7.13\times$ (due to \textbf{DUB} sparsity alone) to $8.5\times$ using combined \textbf{S-DUB} approach.

Interesting to note that, despite the fraction of the entirely removed tiles reducing significantly as the tile size increases, the amount of ADC energy savings achieved by \textbf{S-DUB} method is higher or about the same as the savings achieved by \textbf{DUB} method alone.
Essentially, when there are plenty of crossbars that can be pruned (as in the case of smaller tiles) and the effectiveness of structured pruning is anticipated to be high, \textbf{S-DUB} approach improves results further to account for ADC resolution requirements on the remaining crossbars according to \textbf{DUB} properties.
However, when the effectiveness of structured tile pruning is limited (as in the case of coarse tiles), our method achieves savings in terms of ADC requirements alone by leveraging partially sparse tiles.

\section{Conclusion}
While crossbar-based analog in-memory computing offers an attractive solution for DNN acceleration, the high energy dominance of the peripheral components diminishes its utilization. In this work, we propose a pruning approach inducing \textbf{DUB} sparsity that directly targets improving ADC-specific inefficiencies while maintaining accuracy.
The main idea is to induce sparsity within each crossbar at discretized (\textbf{D}) sparsity levels attuned to the reduction of ADC precision, while simultaneously leveraging unstructured (\textbf{U}) sparsity balanced (\textbf{B}) within neighboring crossbar columns for the amortization of accuracy.
It is achieved by the combination of the training using the newly proposed variance regularization of $L_{0}$ norms of neighboring columns and the pruning of each individual crossbar.
Our experimental results indicate that the proposed approach achieves ADC energy savings by naturally finding the balance between entirely pruning the crossbars and sparsifying the rest of them to reduce ADC precisions.
Its higher energy savings compared to unstructured pruning and structured crossbar pruning methods prove its effectiveness, both as a standalone technique and in combination with structured crossbar compression.

\bibliographystyle{IEEEtran.bst}
\bibliography{main.bib}

\begin{thebibliography}{10}
\providecommand{\url}[1]{#1}
\csname url@samestyle\endcsname
\providecommand{\newblock}{\relax}
\providecommand{\bibinfo}[2]{#2}
\providecommand{\BIBentrySTDinterwordspacing}{\spaceskip=0pt\relax}
\providecommand{\BIBentryALTinterwordstretchfactor}{4}
\providecommand{\BIBentryALTinterwordspacing}{\spaceskip=\fontdimen2\font plus
\BIBentryALTinterwordstretchfactor\fontdimen3\font minus \fontdimen4\font\relax}
\providecommand{\BIBforeignlanguage}[2]{{%
\expandafter\ifx\csname l@#1\endcsname\relax
\typeout{** WARNING: IEEEtran.bst: No hyphenation pattern has been}%
\typeout{** loaded for the language `#1'. Using the pattern for}%
\typeout{** the default language instead.}%
\else
\language=\csname l@#1\endcsname
\fi
#2}}
\providecommand{\BIBdecl}{\relax}
\BIBdecl

\bibitem{simonyan2014very}
K.~Simonyan and A.~Zisserman, ``Very deep convolutional networks for large-scale image recognition,'' \emph{ICLR}, 2015.

\bibitem{he2016deep}
K.~He, X.~Zhang, S.~Ren, and J.~Sun, ``Deep residual learning for image recognition,'' in \emph{IEEE CVPR}, 2016, pp. 770--778.

\bibitem{pmlr-v202-frantar23a}
E.~Frantar and D.~Alistarh, ``{S}parse{GPT}: Massive language models can be accurately pruned in one-shot,'' in \emph{ICML}, vol. 202.\hskip 1em plus 0.5em minus 0.4em\relax PMLR, 23--29 Jul 2023, pp. 10\,323--10\,337.

\bibitem{zong2023detrs}
Z.~Zong, G.~Song, and Y.~Liu, ``Detrs with collaborative hybrid assignments training,'' in \emph{IEEE/CVF ICCV}, 2023, pp. 6748--6758.

\bibitem{han2015deep}
S.~Han, H.~Mao, and W.~J. Dally, ``Deep compression: Compressing deep neural networks with pruning, trained quantization and huffman coding,'' \emph{ICLR}, 2016.

\bibitem{akopyan2015truenorth}
F.~Akopyan, J.~Sawada, A.~Cassidy, R.~Alvarez-Icaza, J.~Arthur, P.~Merolla, N.~Imam, Y.~Nakamura, P.~Datta, G.-J. Nam \emph{et~al.}, ``Truenorth: Design and tool flow of a 65 mw 1 million neuron programmable neurosynaptic chip,'' \emph{IEEE TCAD}, vol.~34, no.~10, pp. 1537--1557, 2015.

\bibitem{shafiee2016isaac}
A.~Shafiee, A.~Nag, N.~Muralimanohar, R.~Balasubramonian, J.~P. Strachan, M.~Hu, R.~S. Williams, and V.~Srikumar, ``Isaac: A convolutional neural network accelerator with in-situ analog arithmetic in crossbars,'' \emph{ACM SIGARCH Computer Architecture News}, vol.~44, no.~3, 2016.

\bibitem{chi2016prime}
P.~Chi, S.~Li, C.~Xu, T.~Zhang, J.~Zhao, Y.~Liu, Y.~Wang, and Y.~Xie, ``Prime: A novel processing-in-memory architecture for neural network computation in reram-based main memory,'' \emph{ACM SIGARCH Computer Architecture News}, vol.~44, no.~3, pp. 27--39, 2016.

\bibitem{song2017pipelayer}
L.~Song, X.~Qian, H.~Li, and Y.~Chen, ``Pipelayer: A pipelined reram-based accelerator for deep learning,'' in \emph{2017 IEEE HPCA}.\hskip 1em plus 0.5em minus 0.4em\relax IEEE, 2017, pp. 541--552.

\bibitem{ankit2019puma}
A.~Ankit, I.~E. Hajj, S.~R. Chalamalasetti, G.~Ndu, M.~Foltin, R.~S. Williams, P.~Faraboschi, W.-m.~W. Hwu, J.~P. Strachan, K.~Roy \emph{et~al.}, ``Puma: A programmable ultra-efficient memristor-based accelerator for machine learning inference,'' in \emph{ASPLOS}, 2019, pp. 715--731.

\bibitem{ankit2020panther}
A.~Ankit, I.~El~Hajj, S.~R. Chalamalasetti, S.~Agarwal, M.~Marinella, M.~Foltin, J.~P. Strachan, D.~Milojicic, W.-M. Hwu, and K.~Roy, ``Panther: A programmable architecture for neural network training harnessing energy-efficient reram,'' \emph{IEEE Transactions on Computers}, vol.~69, no.~8, pp. 1128--1142, 2020.

\bibitem{wan2022compute}
W.~Wan, R.~Kubendran, C.~Schaefer, S.~B. Eryilmaz, W.~Zhang, D.~Wu, S.~Deiss, P.~Raina, H.~Qian, B.~Gao \emph{et~al.}, ``A compute-in-memory chip based on resistive random-access memory,'' \emph{Nature}, vol. 608, no. 7923, pp. 504--512, 2022.

\bibitem{yang2019sparse}
T.-H. Yang, H.-Y. Cheng, C.-L. Yang, I.-C. Tseng, H.-W. Hu, H.-S. Chang, and H.-P. Li, ``Sparse reram engine: Joint exploration of activation and weight sparsity in compressed neural networks,'' in \emph{ISCA}, 2019, pp. 236--249.

\bibitem{chen202115}
Z.~Chen, X.~Chen, and J.~Gu, ``15.3 a 65nm 3t dynamic analog ram-based computing-in-memory macro and cnn accelerator with retention enhancement, adaptive analog sparsity and 44tops/w system energy efficiency,'' in \emph{2021 IEEE ISSCC}, vol.~64.\hskip 1em plus 0.5em minus 0.4em\relax IEEE, 2021, pp. 240--242.

\bibitem{ali202135}
M.~Ali, I.~Chakraborty, U.~Saxena, A.~Agrawal, A.~Ankit, and K.~Roy, ``A 35.5-127.2 tops/w dynamic sparsity-aware reconfigurable-precision compute-in-memory sram macro for machine learning,'' \emph{IEEE Solid-State Circuits Letters}, vol.~4, pp. 129--132, 2021.

\bibitem{kim2023samba}
D.~E. Kim, A.~Ankit, C.~Wang, and K.~Roy, ``Samba: Sparsity aware in-memory computing based machine learning accelerator,'' \emph{IEEE Transactions on Computers}, 2023.

\bibitem{ogbogu2023energy}
C.~Ogbogu, M.~Soumen, B.~K. Joardar, J.~R. Doppa, D.~Heo, K.~Chakrabarty, and P.~P. Pande, ``Energy-efficient reram-based ml training via mixed pruning and reconfigurable adc,'' in \emph{2023 IEEE/ACM ISLPED}.\hskip 1em plus 0.5em minus 0.4em\relax IEEE, 2023, pp. 1--6.

\bibitem{roy2020memory}
K.~Roy, I.~Chakraborty, M.~Ali, A.~Ankit, and A.~Agrawal, ``In-memory computing in emerging memory technologies for machine learning: an overview,'' in \emph{2020 57th ACM/IEEE DAC}.\hskip 1em plus 0.5em minus 0.4em\relax IEEE, 2020, pp. 1--6.

\bibitem{chakraborty2020resistive}
I.~Chakraborty, M.~Ali, A.~Ankit, S.~Jain, S.~Roy, S.~Sridharan, A.~Agrawal, A.~Raghunathan, and K.~Roy, ``Resistive crossbars as approximate hardware building blocks for machine learning: Opportunities and challenges,'' \emph{Proceedings of the IEEE}, vol. 108, no.~12, pp. 2276--2310, 2020.

\bibitem{huang2021mixed}
S.~Huang, A.~Ankit, P.~Silveira, R.~Antunes, S.~R. Chalamalasetti, I.~El~Hajj, D.~E. Kim, G.~Aguiar, P.~Bruel, S.~Serebryakov \emph{et~al.}, ``Mixed precision quantization for reram-based dnn inference accelerators,'' in \emph{2021 26th ASP-DAC}.\hskip 1em plus 0.5em minus 0.4em\relax IEEE, 2021, pp. 372--377.

\bibitem{blalock2020what}
D.~Blalock, J.~J. Gonzalez~Ortiz, J.~Frankle, and J.~Guttag, ``What is the state of neural network pruning?'' in \emph{Proceedings of Machine Learning and Systems}, I.~Dhillon, D.~Papailiopoulos, and V.~Sze, Eds., vol.~2, 2020, pp. 129--146.

\bibitem{han2015learning}
S.~Han, J.~Pool, J.~Tran, and W.~J. Dally, ``Learning both weights and connections for efficient neural network,'' in \emph{NeurIPS}, 2015.

\bibitem{zhang2018systematic}
T.~Zhang, S.~Ye, K.~Zhang, J.~Tang, W.~Wen, M.~Fardad, and Y.~Wang, ``A systematic dnn weight pruning framework using alternating direction method of multipliers,'' in \emph{ECCV}, 2018, pp. 184--199.

\bibitem{rathi2018stdp}
N.~Rathi, P.~Panda, and K.~Roy, ``Stdp-based pruning of connections and weight quantization in spiking neural networks for energy-efficient recognition,'' \emph{IEEE TCAD}, vol.~38, no.~4, pp. 668--677, 2018.

\bibitem{wen2016learning}
W.~Wen, C.~Wu, Y.~Wang, Y.~Chen, and H.~Li, ``Learning structured sparsity in deep neural networks,'' \emph{NeurIPS}, vol.~29, pp. 2074--2082, 2016.

\bibitem{li2016pruning}
H.~Li, A.~Kadav, I.~Durdanovic, H.~Samet, and H.~P. Graf, ``Pruning filters for efficient convnets,'' \emph{ICLR}, 2017.

\bibitem{garg2019low}
I.~Garg, P.~Panda, and K.~Roy, ``A low effort approach to structured cnn design using pca,'' \emph{IEEE Access}, vol.~8, pp. 1347--1360, 2019.

\bibitem{yang2020harmonious}
L.~Yang, Z.~He, and D.~Fan, ``Harmonious coexistence of structured weight pruning and ternarization for deep neural networks,'' in \emph{AAAI}, vol.~34, no.~04, 2020, pp. 6623--6630.

\bibitem{aketi2020gradual}
S.~A. Aketi, S.~Roy, A.~Raghunathan, and K.~Roy, ``Gradual channel pruning while training using feature relevance scores for convolutional neural networks,'' \emph{IEEE Access}, vol.~8, pp. 171\,924--171\,932, 2020.

\bibitem{chu2020pim}
C.~Chu, Y.~Wang, Y.~Zhao, X.~Ma, S.~Ye, Y.~Hong, X.~Liang, Y.~Han, and L.~Jiang, ``Pim-prune: fine-grain dcnn pruning for crossbar-based process-in-memory architecture,'' in \emph{ACM/IEEE DAC}.\hskip 1em plus 0.5em minus 0.4em\relax IEEE, 2020.

\bibitem{liang2018crossbar}
L.~Liang, L.~Deng, Y.~Zeng, X.~Hu, Y.~Ji, X.~Ma, G.~Li, and Y.~Xie, ``Crossbar-aware neural network pruning,'' \emph{IEEE Access}, vol.~6, pp. 58\,324--58\,337, 2018.

\bibitem{lin2019learning}
J.~Lin, Z.~Zhu, Y.~Wang, and Y.~Xie, ``Learning the sparsity for reram: Mapping and pruning sparse neural network for reram based accelerator,'' in \emph{Proceedings of the 24th Asia and South Pacific Design Automation Conference}, 2019, pp. 639--644.

\bibitem{ankit2019trannsformer}
A.~Ankit, T.~Ibrayev, A.~Sengupta, and K.~Roy, ``Trannsformer: Clustered pruning on crossbar-based architectures for energy-efficient neural networks,'' \emph{IEEE Transactions on Computer-Aided Design of Integrated Circuits and Systems}, vol.~39, no.~10, pp. 2361--2374, 2019.

\bibitem{mao2017exploring}
H.~Mao, S.~Han, J.~Pool, W.~Li, X.~Liu, Y.~Wang, and W.~J. Dally, ``Exploring the granularity of sparsity in convolutional neural networks,'' in \emph{Proceedings of the IEEE Conference on Computer Vision and Pattern Recognition Workshops}, 2017, pp. 13--20.

\bibitem{yuan2021tinyadc}
G.~Yuan, P.~Behnam, Y.~Cai, A.~Shafiee, J.~Fu, Z.~Liao, Z.~Li, X.~Ma, J.~Deng, J.~Wang \emph{et~al.}, ``Tinyadc: Peripheral circuit-aware weight pruning framework for mixed-signal dnn accelerators,'' in \emph{2021 DATE}.\hskip 1em plus 0.5em minus 0.4em\relax IEEE, 2021, pp. 926--931.

\bibitem{xue2023hierarchical}
W.~Xue, J.~Bai, S.~Sun, and W.~Kang, ``Hierarchical non-structured pruning for computing-in-memory accelerators with reduced adc resolution requirement,'' in \emph{2023 DATE}.\hskip 1em plus 0.5em minus 0.4em\relax IEEE, 2023, pp. 1--6.

\bibitem{yang2019deephoyer}
H.~Yang, W.~Wen, and H.~Li, ``Deephoyer: Learning sparser neural network with differentiable scale-invariant sparsity measures,'' \emph{ICLR}, 2020.

\bibitem{murmann2020adc}
B.~Murmann, ``Adc performance survey 1997-2020,'' \emph{http://web.stanford.edu/\~{}murmann/adcsurvey.html}, 2020.

\bibitem{krizhevsky2009learning}
A.~Krizhevsky, G.~Hinton \emph{et~al.}, ``Learning multiple layers of features from tiny images,'' University of Toronto, Toronto, ON, Canada, Tech. Rep., 2009.

\bibitem{russakovsky2015imagenet}
O.~Russakovsky, J.~Deng, H.~Su, J.~Krause, S.~Satheesh, S.~Ma, Z.~Huang, A.~Karpathy, A.~Khosla, M.~Bernstein \emph{et~al.}, ``Imagenet large scale visual recognition challenge,'' \emph{International journal of computer vision}, vol. 115, pp. 211--252, 2015.

\bibitem{yuan2006model}
M.~Yuan and Y.~Lin, ``Model selection and estimation in regression with grouped variables,'' \emph{Journal of the Royal Statistical Society: Series B (Statistical Methodology)}, vol.~68, no.~1, pp. 49--67, 2006.

\end{thebibliography}

\begin{IEEEbiography}[{\includegraphics[width=1in,height=1.25in,clip,keepaspectratio]{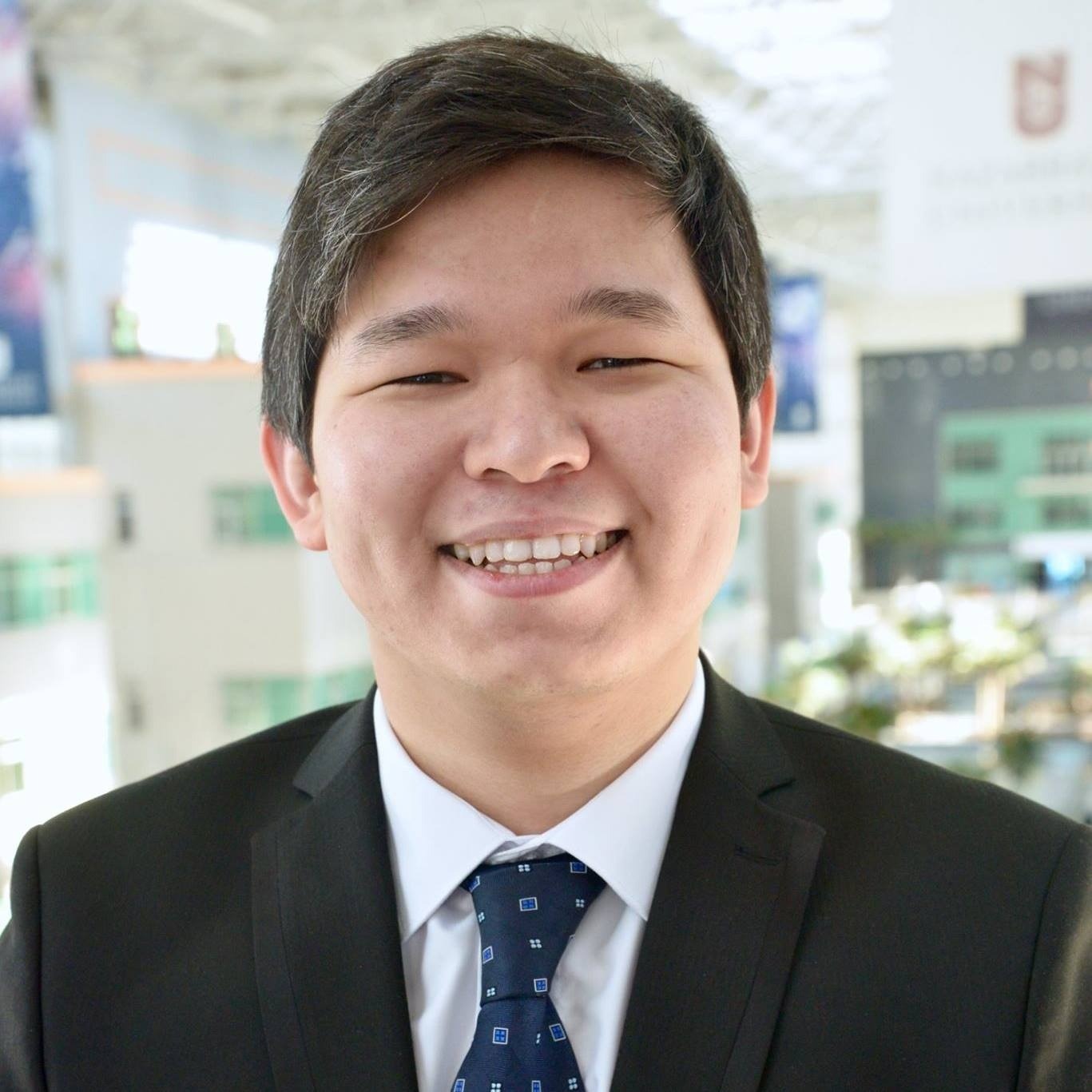}}]
{Timur Ibrayev} received the B.S. in electrical and electronics engineering from Nazarbayev University, Astana, Kazakhstan, in 2017. He is currently pursuing the Ph.D. degree in electrical and computer engineering at Purdue University, West Lafayette, IN, USA. He interned at Intel Corp. during Summer 2022, where he worked on profiling and optimization of the deep learning code base for their clusters based on Gaudi deep learning processors. His research interests include hardware-software co-design and neuro-inspired computer vision algorithms.
\end{IEEEbiography}

\begin{IEEEbiography}[{\includegraphics[width=1in,height=1.25in,clip,keepaspectratio]{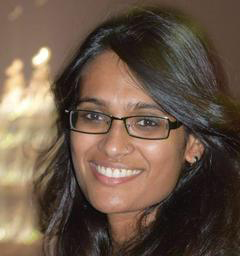}}]{Isha Garg} received the B.Tech. degree in electrical engineering from BITS Pilani, India, in 2013, and the Ph.D. degree from Purdue University, in 2023, under the guidance of Prof. Kaushik Roy. She has worked at AMD and Synopsys on memory design, at National University of Singapore, researching efficient inference of convolutional neural networks with Prof. Massimo Alioto, and is currently at Apple working in the video engineering team. Her research interests include efficient, robust, and private training and inference algorithms for deep learning and computer vision.
\end{IEEEbiography}

\begin{IEEEbiography}[{\includegraphics[width=1in,height=1.25in,clip,keepaspectratio]{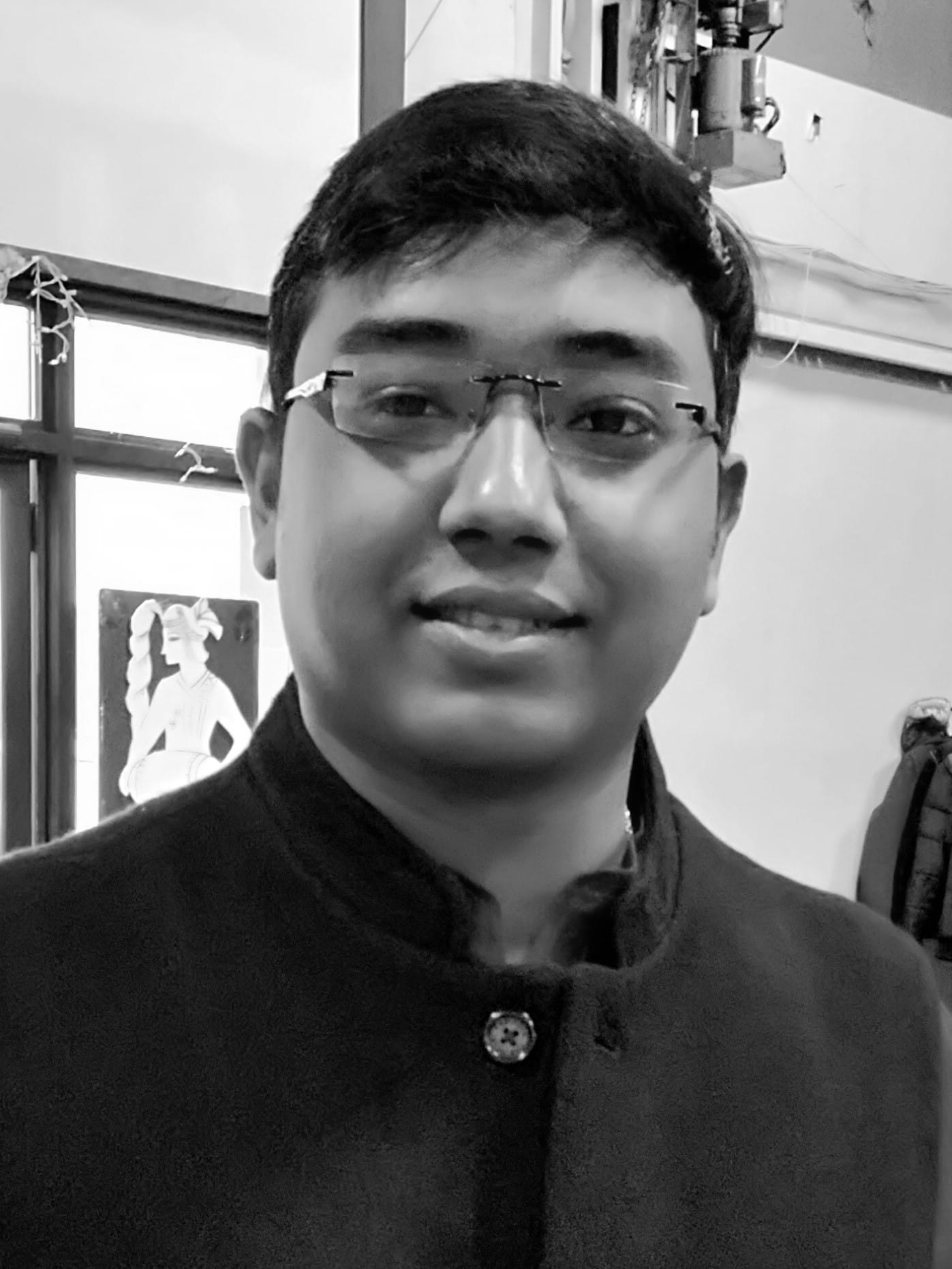}}]{Indranil Chakraborty} is currently a Hardware Engineer at Google, Sunnyvale, California. He received the B.Engg. degree in electronics and telecommunication engineering from Jadavpur University, Kolkata, India, in 2013, an M.Tech. degree in electrical engineering from Indian Institute of Technology Bombay, Mumbai, India, in 2016, and the Ph.D. degree in 2021 at Nanoelectronics Research Laboratory, Purdue University, West Lafayette, IN, USA. His primary research interests lie in architecture and design of hardware accelerators for machine-learning workloads using CMOS and emerging technologies.
\end{IEEEbiography}

\begin{IEEEbiography}[{\includegraphics[width=1in,height=1.25in,clip,keepaspectratio]{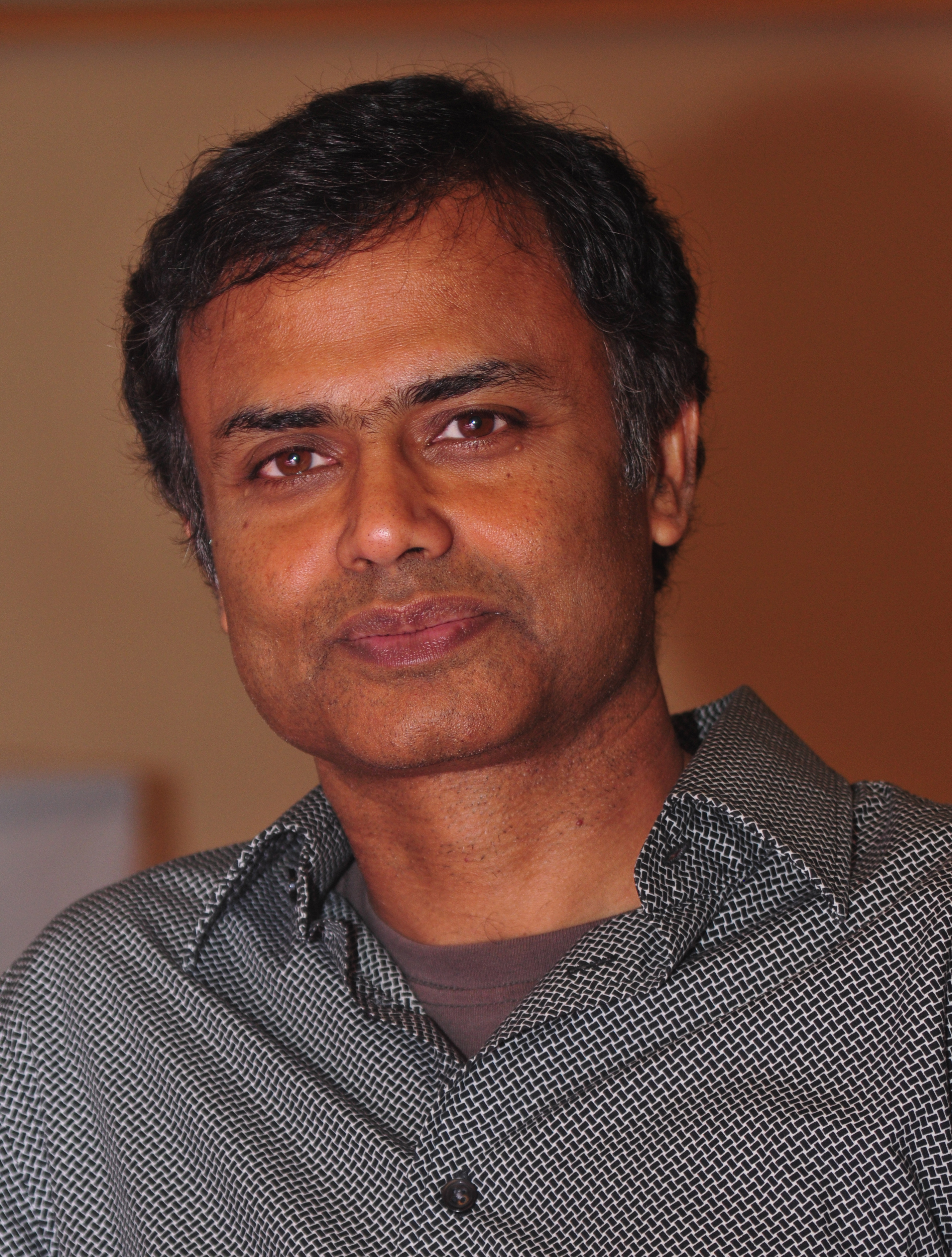}}]{Kaushik Roy (Fellow, IEEE)}  received the B.Tech. degree in electronics and electrical communications engineering from IIT Kharagpur, Kharagpur, India, and the Ph.D. degree from the electrical and computer engineering department, University of Illinois at Urbana–Champaign, in 1990.

He was with the Semiconductor Process and Design Center of Texas Instruments, Dallas, where
he worked on FPGA architecture development and low-power circuit design. He joined the Electrical and Computer Engineering Faculty, Purdue University, West Lafayette, IN, USA, in 1993, where he is
currently the Edward G. Tiedemann Jr. Distinguished Professor. He is the Director of the Center for Brain-Inspired Computing (C-BRIC) funded by SRC/DARPA. He has published more than 700 articles in refereed journals and conferences, holds 25 patents, supervised more than 100 Ph.D. dissertations, and is co-author of two books on Low Power CMOS VLSI Design (John Wiley \& McGraw Hill). His research interests include neuromorphic and emerging computing models, neuro-mimetic devices, spintronics, device-circuitalgorithm co-design for nano-scale silicon, non-silicon technologies, and low-power electronics. He received the National Science Foundation Career Development Award, in 1995, the IBM Faculty Partnership Award, the ATT/Lucent Foundation Award, the 2005 SRC Technical Excellence Award,
the SRC Inventors Award, the Purdue College of Engineering Research Excellence Award, the Humboldt Research Award, in 2010, the 2010 IEEE Circuits and Systems Society Technical Achievement Award (Charles Desoer Award), the Distinguished Alumnus Award from IIT Kharagpur, the Fulbright-Nehru Distinguished Chair, the DoD Vannevar Bush Faculty Fellow, from 2014 to 2019, the Semiconductor Research Corporation Aristotle Award, in 2015, the 2020 Arden Bement Jr. Award, the Highest Research
Award given by Purdue University in pure and applied science and engineering, and the Best Paper Awards at 1997 International Test Conference, the IEEE 2000 International Symposium on Quality of IC Design, the 2003 IEEE Latin American Test Workshop, the 2003 IEEE Nano, the 2004 IEEE International Conference on Computer Design, the 2006 IEEE/ACM International Symposium on Low Power Electronics and Design, and the 2005 IEEE Circuits and System Society Outstanding Young Author Award (Chris Kim), the 2006 IEEE TRANSACTIONS ON VLSI SYSTEMS Best Paper Award, the 2012 ACM/IEEE International Symposium on Low Power Electronics and Design Best Paper Award, and the 2013 IEEE TRANSACTIONS ON VLSI Best Paper Award. He was a Purdue University Faculty Scholar from 1998 to 2003. He was a Research Visionary Board Member of Motorola Labs, in 2002 and held the M. Gandhi Distinguished Visiting Faculty with IIT Bombay and the Global Foundries Visiting Chair with the National University of Singapore. He has been in the editorial board of IEEE Design and Test, the IEEE TRANSACTIONS ON CIRCUITS AND SYSTEMS, the IEEE TRANSACTIONS ON
VLSI SYSTEMS, and the IEEE TRANSACTIONS ON ELECTRON DEVICES. He was a Guest Editor for Special Issue on Low-Power VLSI in IEEE Design and Test, in 1994, the IEEE TRANSACTIONS ON VLSI SYSTEMS, in June 2000, the IEEE Proceedings–Computers and Digital Techniques, in July 2002, and the IEEE JOURNAL ON EMERGING AND SELECTED TOPICS IN CIRCUITS AND SYSTEMS, in 2011. 
\end{IEEEbiography}

\end{document}